# Network Slicing Games: Enabling Customization in Multi-Tenant Networks

Pablo Caballero*    Albert Banchs†    Gustavo de Veciana*    Xavier Costa-Pérez‡
*The University of Texas at Austin, Austin, TX. Email: pablo.caballero@utexas.edu, Gustavo@ece.utexas.edu
†University Carlos III of Madrid and IMDEA Networks Institute, Madrid. Spain. Email: banchs@it.uc3m.es
‡NEC Laboratories Europe, Heidelberg, Germany. Email: xavier.costa@neclab.eu

*Abstract*—Network slicing to enable resource sharing among multiple tenants –network operators and/or services– is considered a key functionality for next generation mobile networks. This paper provides an analysis of a well-known model for resource sharing, the 'share-constrained proportional allocation' mechanism, to realize network slicing. This mechanism enables tenants to reap the performance benefits of sharing, while retaining the ability to customize their own users' allocation. This results in a *network slicing game* in which each tenant reacts to the user allocations of the other tenants so as to maximize its own utility. We show that, under appropriate conditions, the game associated with such strategic behavior converges to a Nash equilibrium. At the Nash equilibrium, a tenant always achieves the same, or better, performance than under a static partitioning of resources, hence providing the same level of protection as such static partitioning. We further analyze the efficiency and fairness of the resulting allocations, providing tight bounds for the price of anarchy and envy-freeness. Our analysis and extensive simulation results confirm that the mechanism provides a comprehensive practical solution to realize network slicing. Our theoretical results also fill a gap in the literature regarding the analysis of this resource allocation model under strategic players.

## I. INTRODUCTION

There is consensus among the relevant industry and standardization communities [1], [2] that a key element in 5G mobile networks will be *network slicing*. The idea is to allow the mobile infrastructure to be "sliced" into logical networks, which are operated by different entities and may be tailored to support specific services. This provides a basis for efficient infrastructure sharing among diverse entities, ranging from classical or virtual mobile network operators to new players that simply view connectivity as a service. Such new players could be, for instance, Over-The-Top (OTT) service providers which use a *network slice* to ensure satisfactory service to their customers (e.g., Amazon Kindle's support for downloading content or a pay TV channel including a premium subscription). In the literature, the term *tenant* is often used to refer to the owner of a network slice.

A network slice is a collection of resources and functions that are orchestrated to support a specific service. This includes software modules running at different locations as well as the nodes' computational resources, and communication resources in the backhaul and radio network. The intention is to only provide what is necessary for the service, avoiding unnecessary overheads and complexity. Thus, network slices enable tenants to compete with each other using the same physical infrastructure, but customizing their slices and network operation according to their market segment's characteristics and requirements. For instance, slices can be geared at supporting various IoT or M2M applications, such as the connectivity required to realize 'intelligent' vehicular systems.

A key problem underlying network slicing is enabling efficient sharing of mobile network resources. One of the approaches considered in 3GPP suggests that resources could be statically partitioned based on fixed 'network shares' [3]. However, given that slices' loads may be spatially inhomogenous and time varying, it is desirable to allow resource allocations to be 'elastic', e.g., dependent on the slices' loads at different base stations. At the same time, tenants should be protected from one another, and retain the ability to autonomously manage their slice's resources, in order to better customize allocations to their customers. To that end, it is desirable to adopt resource allocation models in which tenants can communicate their preferences to the infrastructure (say by dynamically subdividing their network share amongst their customers) and then have base stations' resources allocated according to their preferences (i.e., proportionally to the customers' shares).

Under such a dynamic resource allocation model, a tenant might exhibit strategic behavior, by adjusting its preferences depending on perceived congestion at resources, so as to maximize its own utility. Such behavior could in turn have adverse effects on the network; for instance, the overall efficiency may be harmed, or one may see instability in slice requests. The focus of this paper is on ($i$) the analysis and performance of this simple resource allocation model, and ($ii$) the validation of its feasibility as a means to enable tenants to customize resource allocation within their slice while protecting them from one another.

*Related work:* The resource allocation mechanism informally described above corresponds to a Fisher market, which is a standard framework in economics. In such markets, buyers (in our case slices) have fixed budgets (in our case network shares) and (according to their preferences) bid for resources within their budget, which are then allocated to buyers proportionally to their bids. Analysis of the Fisher market shows that, as long as buyers are price-taking (i.e., they do not anticipate the impact of their bids on the price – in our case, the impact of the slices' preferences on the overall congestion), the Nash equilibrium is socially optimal, and distributed algorithms can be easily devised to reach it [4]. This assumption may be reasonable for markets where the impact of a single buyer on a resource's price is negligible, but does not apply to our case where a relatively small number of active tenants might be sharing resources.

There is a substantial literature on Fisher markets with strategic buyers, which, as will be studied in this paper, anticipate the impact of their bids [5]. The analysis, so far, has been limited to the case of buyers with *linear utility* functions of the allocated resources, which can lead to extremely unfair allocations. While such utility functions may be suitable for goods, they are not an appropriate model for tenants wishing to customize allocations amongst their customers. This paper includes a comprehensive analysis for a wide set of slice utility functions, including the convergence of best response dynamics and other results which to our knowledge are new.

A related resource allocation model often considered in the networking field is the so-called 'Kelly's mechanism' [6]; this mechanism allocates resources to players proportionally to their bids and, assuming that they are price-taking, converges to a social optimum. Follow-up work has considered price-anticipating players in this setting; for example, [7] analyze efficiency losses, while [8] devise a scalar-parametrized modification that is once again socially optimal for price-anticipating players. However, in Kelly's mechanism players respond to their payoff (given by the utility minus cost) whereas in our model tenants' behavior is only driven by their utilities (since they have a fixed budget: the network share). Consequently, results on the analysis of Kelly's mechanism are not applicable to our setting.

Table I capture the main resource allocation models for this problem highlighting some of the most relevant work for each problem and situating the contribution of this work.

|  | price taking | price anticipating | |
| --- | --- | --- | --- |
|  | scalar bid | scalar bid | vector bid |
| non fixed budget | [6] Kelly's mechanism (conv, efficiency) | VCG-Kelly mechanism [8] Hajek/Yang [9] Johari/Tsitsiklis | Johari/Tsitsiklis [7] Efficiency of congestion games |
|  | concave utilities | linear utilities | concave utilities |
| fixed budget | [4] Zhang (convergence) | [5] Zhang (conv, efficiency) | *This work* (conv, efficiency) |

TABLE I: Resource allocation models.

From a more practical angle, multi-tenant sharing has been studied from different points of view, including planning, economics, coverage, performance, etc. [10], [11]. This paper focuses specifically on the design of algorithms for resource sharing among tenants, which has been previously addressed by [12]–[15]. The work of [15] considers sharing via a bid-based auction, which may incur substantial overhead and complexity; in contrast, our approach relies on fixed (pre-negotiated) network shares. The works of [12]–[14] also fix a network share per slice, but consider approaches where the infrastructure makes centralized decisions on the resources allocated to each tenant's customers; hence, these approaches do not enable tenants to make their own decisions on how to allocate resources to their customers.

Network slicing has emerged as a desirable feature for 5G [1]. 3GPP has started work on defining requirements for network slicing [2], whereas the Next Generation Mobile Network (NGMN) alliance has identified network sharing among slices (the focus of this paper) as a key issue [16]. In spite of these efforts, most of the work so far has addressed architectural aspects with only a limited focus on resource allocation algorithms [17], [18]. To the best of our knowledge, this is the first work investigating how to enable tenants to customize their allocations in a dynamic slicing model.

*Key contributions:* The rest of the paper is organized as follows. After introducing our system model (Section II), we show that with the resource sharing model under study, each slice has the ability to achieve the same or better utility than under static resource slicing irrespective of how the other slices behave, which confirms that this model effectively protects slices from one another (Section III-A). Next we show that if tenants exhibit strategic behavior (i.e, optimize their utilities), then ($i$) a Nash equilibrium exists under mild conditions; and ($ii$) the system converges to such an equilibrium when tenants sequentially take their best response (Sections III-B and III-C). The resulting efficiency and fairness among tenants are then studied, providing: ($i$) a tight bound on the Price of Anarchy of the system, and ($ii$) a bound on the Envy-freeness (Section IV). Our results are validated via simulation, confirming that the approach provides substantial gains, protects network slices from each other, operates close to optimal performance and is effectively envy-free (Section V).

## II. SYSTEM MODEL

We consider a wireless network consisting of a set of resources $\mathcal{B}$ (the base stations or sectors) shared by a set of network slices $\mathcal{O}$ (the tenants). At a given point in time, the network supports a set of users $\mathcal{U}$ (the customers or devices), which can be subdivided into subsets $\mathcal{U}_b$ (the users at base station $b$), $\mathcal{U}^o$ (the users of slice $o$) and $\mathcal{U}_b^o$ (their intersection). We further assume that a user $u \in \mathcal{U}$ has a mean peak capacity $c_u$ depending on the choice of modulation and coding at the base station it is associated with. For any user $u$, we let $b(u)$ denote the base station it is currently associated with.

### A. Resource allocation model

As indicated in the introduction, we focus on a well established resource sharing model known in economics as a Fisher market. Hereafter, we will refer to this model as the 'Share-Constrained Proportional Allocation' (SCPA) mechanism.

In our setting, each slice $o$ is allocated a network share $s_o$ (corresponding to its budget) such that $\sum_{o \in \mathcal{O}} s_o = 1$. The slice is at liberty in turn to distribute its share amongst its users, assigning them weights (corresponding to the bids): $w_u$ for $u \in \mathcal{U}^o$, such that $\sum_{u \in \mathcal{U}_o} w_u = s_o$. We let $\mathbf{w}^o = (w_u : u \in \mathcal{U}_o)$ be the weights of slice $o$, $\mathbf{w} = (w_u : u \in \mathcal{U})$ those of all slices and $\mathbf{w}^{-o} = (w_u : u \in \mathcal{U} \setminus \mathcal{U}^o)$ the weights of all users excluding those of slice $o$.

We shall assume users are allocated a fraction of resources at their base station proportionally to their weights $w_u$. Thus the rate of user $u$ is given by

$$r_u(\mathbf{w}) = \frac{w_u}{\sum_{v \in \mathcal{U}_{b(u)}} w_v} c_u = \frac{w_u}{l_{b(u)}(\mathbf{w})} c_u$$

where $l_b(\mathbf{w}) = \sum_{u \in \mathcal{U}_b} w_u$ denotes the overall load at $b$ (recall that $c_u$ is the achievable rate if the user had the entire base station to itself).

To implement the above resource allocation, a slice needs to communicate the weights of its users $\mathbf{w}^o$ to the infrastructure. When selecting its weights, we assume that the slice is aware of the overall load at each base station (indeed, a slice could infer these by varying its users' weights and observing the resulting resource allocations).[1]

In the case where a slice $o$ is the only one with users at a given base station $b$, we shall assume that the slice's users are allocated the entire capacity at that base station independent of their weights. Thus such a slice would set $w_u = 0$ for these users, allowing them to receive all the resources of this base station without consuming any share. In order to avoid dealing with this special case, and without loss of generality, we will make the following assumption for the rest of the paper.

**Assumption 1.** *(Competition at all resources) We assume that all resources have active users from at least two slices.*

### B. Network Slice Utility and Service Differentiation

Network slices may support services and customers of different types and needs. Alternatively, competing slices with similar customer types may wish to differentiate the service they provide. To that end, we assume each network slice has a *private* utility that reflects the benefit obtained by the slice from a given allocation and is given by

$$U^o(\mathbf{w}) = \sum_{u \in \mathcal{U}^o} \phi_u f_u(r_u(\mathbf{w})),$$

where $\phi_u$ is the relative priority of user $u$, with $\phi_u \geq 0$ and $\sum_{u \in \mathcal{U}_o} \phi_u = 1$, and $f_u(\cdot)$ is a (concave) utility function associated with the user. In the sequel, we will often focus on the following well-known class of utility functions [19].

**Definition 1.** *A network slice $o$ has a homogenous $\alpha_o$-fair utility if for all $u \in \mathcal{U}^o$ we have that*

$$f_u(r_u) = \begin{cases} \frac{(r_u)^{1-\alpha_o}}{(1-\alpha_o)}, & \alpha_o \neq 1 \\ \log(r_u), & \alpha_o = 1. \end{cases}$$

Thus, in our setting, a slice is free to choose different fairness criteria in allocating resources across its users, by selecting the appropriate $\alpha_o$ parameter. Note that $\alpha_o = 1$ corresponds to the widely accepted proportional fairness criterion, while $\alpha_o = 2$ corresponds to potential delay fairness, $\alpha_o \to \infty$ to max-min fairness and $\alpha_o = 0$ to linear sum utility.

---
[1]It is worth noting that, with the SCPA mechanism under study, the weights of a given tenant are not disclosed to the others, which only see the overall load at each base station.

A slice can also 'strategically' optimize the weight allocation of its users to maximize its own utility. We will consider such strategic behavior of weight allocations in Section III.

### C. Baseline allocations

Next we introduce two natural resource allocation comparative baselines: socially optimal allocations and static slicing.

*a) Socially Optimal Allocations (SO):* If slices were to share their utility functions with a centralized authority, one could in principle consider a socially optimal allocation of weights and resources. These would be given by the maximizer to the *overall network utility* $U(\mathbf{w})$ given by (see [14]):

$$\max_{\mathbf{w} \geq 0} \quad U(\mathbf{w}) := \sum_{o \in \mathcal{O}} s_o U^o(\mathbf{w})$$

$$\text{s.t.} \quad r_u(\mathbf{w}) = \frac{w_u}{l_{b(u)}(\mathbf{w})} c_u, \quad \forall u \in \mathcal{U}$$

$$\sum_{u \in \mathcal{U}^o} w_u = s_o, \quad \forall o \in \mathcal{O}.$$

Note that (as in [14]) we have weighted the slices' utilities to reflect their shares (thus prioritizing those with higher shares). We shall denote the resulting optimal weight and resource allocations under the socially optimal allocations by $\mathbf{w}^*$ and $\mathbf{r}^* = (r_u^* : u \in \mathcal{U})$, respectively.

*b) Static Slicing (SS):* By static slicing (also known as static splitting [20]) we refer to a complete partitioning of resources based on the network shares $s_o, o \in \mathcal{O}$. In this setting, each slice $o$ receives a fixed fraction $s_o$ of each resource and can unilaterally optimize its weight allocation as follows:

$$\max_{\mathbf{w}^o \geq 0} \quad U^o(\mathbf{w}^o) = \sum_{u \in \mathcal{U}^o} \phi_u f_u(r_u(\mathbf{w}^o))$$

$$\text{s.t.} \quad r_u(\mathbf{w}^o) = \frac{w_u}{\sum_{v \in \mathcal{U}_{b(u)}^o} w_v} s_o c_u \quad \forall u \in \mathcal{U}^o$$

$$\sum_{u \in \mathcal{U}^o} w_u = s_o,$$

where we have abused notation to indicate that, in this case, $U^o$ and $r_u$ depend only on $\mathbf{w}^o$. We shall denote the resulting optimal weight and resource allocations under static slicing for all slices by $\mathbf{w}^{ss}$ and $\mathbf{r}^{ss} = (r_u^{ss} : u \in \mathcal{U})$ respectively, where

$$r_u^{ss} = \frac{w_u^{ss}}{\sum_{v \in \mathcal{U}_{b(u)}^o} w_v^{ss}} s_o c_u \quad \forall u \in \mathcal{U}^o, \forall o \in \mathcal{O}. \quad (1)$$

### III. STRATEGIC BEHAVIOR AND NASH EQUILIBRIUM

Under the SCPA resource allocation model, it is reasonable to assume that a player (network slice) would choose to adjust its weights so as to optimize its utility (and thus the service delivered to its customers). Since the resources allocated to a user depend on the weight allocations of the other slices, such behavior would be predicated on the aggregate weight of the other slices at each resource. From the point of view of slice $o$, the overall load at resource $b$ can be decomposed as

$$l_b(\mathbf{w}) = a_b^o(\mathbf{w}^{-o}) + d_b^o(\mathbf{w}^o)$$

where

$$a_b^o(\mathbf{w}^{-o}) = \sum_{o' \in \mathcal{O} \setminus \{o\}} \sum_{u \in \mathcal{U}_b^{o'}} w_u \quad \text{and} \quad d_b^o(\mathbf{w}^o) = \sum_{u \in \mathcal{U}_b^o} w_u,$$

correspond to the aggregate weight of the other slices and that of slice $o$, respectively. As indicated in Section II, we assume $\boldsymbol{a}^o(\mathbf{w}^{-o}) = (a_b^o(\mathbf{w}^{-o}) : b \in \mathcal{B})$ are readily available to slice $o$.

### A. Gain over Static Slicing

We first analyze if strategic behavior on the part of network slices may result in allocations that are worse that those under static slicing. Note that static slicing provides complete isolation among slices but potentially poor utilization. A critical question is whether dynamic sharing, which achieves a higher resource utilization, also provides the same level of protection. This is confirmed by the following result.

**Lemma 1.** *Consider slice $o$ and any feasible weight allocation $\mathbf{w}^{-o}$ for other slices satisfying the network share constraints. Then, there exists a weight allocation $\mathbf{w}^o$ for slice $o$, possibly dependent on $\mathbf{w}^{-o}$, such that the resulting weight allocation $\mathbf{w}$ satisfies $r_u(\mathbf{w}) \geq r_u^{ss}$ for all $u \in \mathcal{U}_o$.*

This lemma is easily shown by choosing $\mathbf{w}^o$ such that

$$w_u = \frac{w_u^{ss}}{\sum_{u \in \mathcal{U}_{b(u)}^o} w_u^{ss}} \frac{a_{b(u)}^o(\mathbf{w}^{-o})}{\sum_{b' \in \mathcal{B}_o} a_{b'}^o(\mathbf{w}^{-o})} s_o, \quad \forall u \in \mathcal{U}^o$$

where $\mathcal{B}^o$ is the set of base stations where slice $o$ has users. The intuitive interpretation for this choice is that by distributing its weights proportionally to the load at each base station, slice $o$ can achieve the same resource allocation as static slicing at each base station. Further, by redistributing these allocations amongst its user in the same manner as static slicing, it achieves at least as much rate per user.

It follows immediately from this lemma that under the SCPA resource allocation model, if all slices exhibit strategic behavior attempting to maximize their utilities, they necessarily achieve a higher utility than under static slicing.

**Theorem 1.** *If the game where each network slice maximizes its utility has a Nash equilibrium, then each slice achieves a higher utility than under static slicing.*

Note this result does not require slices to have homogenous or concave utilities, just that they be increasing in the users' rate allocations.

### B. Existence of Nash Equilibrium

Next we study whether there exists a Nash equilibrium (NE) under which no slice can benefit by unilaterally changing its weight allocation. To that end, we first characterize the best response of a slice.

Given the weights of the other slices, $\mathbf{w}^{-o}$, the best response of slice $o$ is the unique maximizer $\mathbf{w}^o$ of its utility, i.e.,

$$\max_{\mathbf{w}'^o \geq 0} \sum_{u \in \mathcal{U}_o} \phi_u f_u \left( \frac{w_u' c_u}{a_{b(u)}^o(\mathbf{w}^{-o}) + d_{b(u)}^o(\mathbf{w}'^o)} \right)$$

$$\text{s.t} \sum_{u \in \mathcal{U}_o} w_u' = s_o.$$

The following lemma characterizes the best response for a network slice with homogenous $\alpha_o$-fair utility (see [5] for the best response when $\alpha_o = 0$).

**Lemma 2.** *Suppose slice $o$ has a homogeneous $\alpha_o$-fair utility (with $\alpha_o > 0$). Given the weights of the other slices $\mathbf{w}^{-o} > 0$, slice $o$'s best response $\mathbf{w}^o$ is the unique solution to the following nonlinear set of equations:*

$$w_u = \frac{\beta_u \dfrac{\left(a_{b(u)}^o(\mathbf{w}^{-o})\right)^{\frac{1}{\alpha_o}}}{\left(a_{b(u)}^o(\mathbf{w}^{-o}) + d_{b(u)}^o(\mathbf{w}^o)\right)^{\frac{2}{\alpha_o} - 1}}}{\displaystyle\sum_{v \in \mathcal{U}_o} \beta_v \dfrac{\left(a_{b(v)}^o(\mathbf{w}^{-o})\right)^{\frac{1}{\alpha_o}}}{\left(a_{b(v)}^o(\mathbf{w}^{-o}) + d_{b(v)}^o(\mathbf{w}^o)\right)^{\frac{2}{\alpha_o} - 1}}} s_o, \quad \forall u \in \mathcal{U}^o, \quad (2)$$

*where $\beta_u := (\phi_u)^{\frac{1}{\alpha_o}} (c_u)^{\frac{1}{\alpha_o} - 1}$.*

Note that slice $o$ need only know $\boldsymbol{a}^o(\mathbf{w}^{-o})$ to compute its best response. Building on this characterization, we will study the game in which all slices choose to allocate their weights based on their best response. The following theorem proves that this game admits a Nash equilibrium, i.e., there is a weight allocation $\mathbf{w}$ such that no slice can improve its utility by modifying its weights unilaterally.[2]

**Theorem 2.** *Suppose all slices have homogenous $\alpha_o$-fair utilities (with possibly different $\alpha_o > 0$). Then, there exists a (not necessarily unique) Nash equilibrium satisfying (2) for each slice.*

The proof of this result is technical and been relegated to the Appendix. The argument proceeds as follows. We consider a perturbed game where an additional slice assigns a weight $\varepsilon$ at each base station. For this perturbed game, we have concave utilities and compact strategy spaces such that the result of [21] gives existence of a Nash equilibrium.[3] We then consider a sequence of such equilibria as $\varepsilon \to 0$. By compactness of the strategy space it must have a converging subsequence. One can further show that the weight allocations for the perturbed equilibria have uniform positive lower bounds, so the limit of the converging subsequence also has positive weights. Note that (as it can be seen from Lemma 2) slice $o$'s best response in the perturbed game $\mathbf{w}^o$ is a continuous differentiable function of $\mathbf{w}^{-o}$ as long as $\mathbf{w}^{-o} > 0$. It then follows by continuity that the limit of the converging subsequence is a Nash equilibrium.

---

[2]The existence of a NE had already been proven by [4] for the case $\alpha_o = 0 \; \forall o$. Here we extend this result to any combination of $\alpha_o$ values.

[3]In particular, [21] shows by applying the Kakutani fixed point theorem that there exists a solution to the equations defining a Nash equilibrium. Note that in the case of users with log utilities, the function is not defined for a weight of 0 and hence the conditions of [21] are not satisfied; however, a careful reading of the proof of [21] shows the result still applies.

## C. Convergence of Best Response Dynamics

Below we will consider a best response game wherein slices realize their best responses in rounds; specifically, they update their weights ($\mathbf{w}^o$) sequentially, one at a time and in the same fixed order, in response to the other slices' weights ($\mathbf{a}^o$).

**Theorem 3.** *If slices have homogeneous $\alpha_o$-fair utilities, possibly with different $\alpha_o \in [1,2]$ for $o \in \mathcal{O}$, then the best response game converges to a Nash equilibrium.*

Note that the value of $\alpha_o$ impacts a slice's best response and consequently the game dynamics. As seen in Lemma 2, the best response weights are proportional to:

$$w_u \propto g(a_b^o, d_b^o) := \frac{(a_b^o)^{\frac{1}{\alpha_o}}}{(a_b^o + d_b^o)^{\frac{2}{\alpha_o}-1}},$$

where we have suppressed the dependency of $a_b^o$ on $\mathbf{w}^{-o}$ and $d_b^o$ on $\mathbf{w}^o$. The function $g(\cdot,\cdot)$ has different properties depending on $\alpha_o$ which are shown in Table II. The regime where $1 \leq \alpha_o \leq 2$, considered in Theorem 3, is of particular interest since it includes proportional ($\alpha_o = 1$) and potential delay ($\alpha_o = 2$) fairness. It is known that convergence is not ensured when $\alpha_o = 0$ for all slices (see [5]); for other regimes, we resort to the simulations results of Section V, which suggest convergence for any $\alpha_o > 0$.

|  | $\alpha_o = 0$ | $0 < \alpha_o < 1$ | $1 \leq \alpha_o \leq 2$ | $2 < \alpha_o < \infty$ |
|---|---|---|---|---|
| $g$ w.r.t. $d_b^o$ | linear | convex | convex | concave |
| $g$ w.r.t. $a_b^o$ | linear | convex | concave | concave |
| NE existence | ✓ [5] | ✓Theorem 2 for heterogeneous $\alpha_o$ | | |
| convergence | ✗ [5] | ✓simulations | ✓Theorem 3 | ✓simulations |

TABLE II: Impact of $\alpha_o$ on slice's Best Responses.

Perhaps surprisingly, the above result is quite challenging to show. The key challenge lies in the "price-anticipating" aspect of the best response, in which players anticipate the impact of their own allocation.[4] The rest of this section is a sketch of the proof for this result.

We shall denote time as slotted $\{0, 1, ..., t, ...\}$ and assume a single slice makes an update each time slot. Without loss of generality, we will index slices $\{1, 2, ..., |\mathcal{O}|\} = \mathcal{O}$ according to their updating order in a round. We let $\mathbf{w}(t) = (\mathbf{w}^o(t) \colon o \in \mathcal{O})$ be the weights of all slices at the end the time slot $t$ update, where $\mathbf{w}^o(t) = (w_u(t) \colon u \in \mathcal{U}^o)$. Suppose that slices have arbitrary positive initial weight vectors at time zero denoted $\mathbf{w}(0) = (\mathbf{w}^1(0), \mathbf{w}^2(0), ..., \mathbf{w}^{|\mathcal{O}|}(0))$. Consequently, slice 1 will update its weights at time slots: $\{1, |\mathcal{O}|+1, ..., r \cdot |\mathcal{O}|+1\}$, corresponding to rounds $\{0, 1, ..., r, ...\}$.

We will further define $\mathbf{\Delta w}^o(t+1) = (\Delta w_u(t+1) \colon u \in \mathcal{U}^o)$, where $\mathbf{\Delta w}^o(t+1) = (\Delta w_u(t+1) \colon u \in \mathcal{U}^o)$ such that,

$$w_u(t+1) = w_u(t)(1 + \Delta w_u(t+1)), \quad \forall o \in \mathcal{O}, u \in \mathcal{U}^o$$

---
[4]Indeed, as mentioned in the introduction, there are very few results in the literature on the convergence of price-anticipating best response dynamics.

where $1 + \Delta w_u(t+1)$ captures the relative change in slice $o$'s weight update at time slot $t+1$. Furthermore, to capture the overall changes in slices weights at the end of each round, we shall define $\boldsymbol{\omega}(0) = \mathbf{w}(0)$, $\boldsymbol{\omega}(r) = (\boldsymbol{\omega}^o(r) \colon o \in \mathcal{O})$ where $\boldsymbol{\omega}^o(r) = \mathbf{w}^o(r \cdot |\mathcal{O}|+1)$ and $\mathbf{\Delta \boldsymbol{\omega}}^o(r)$ such that $\mathbf{\Delta \boldsymbol{\omega}}(r) = (\Delta \omega_u^o(r) \colon u \in \mathcal{U}^o)$. For all $o \in \mathcal{O}$, we define

$$\overline{\Delta}\omega^o(r) := \max_{u \in \mathcal{U}^o} \Delta \omega_u^o(r), \quad \underline{\Delta}\omega^o(r) := \min_{u \in \mathcal{U}^o} \Delta \omega_u^o(r).$$

The key step in our convergence proof is the following lemma – see the appendix for a detailed proof.

**Lemma 3.** *If the game has not converged to a Nash equilibrium, i.e. $\mathbf{\Delta \boldsymbol{\omega}}(r) \neq \mathbf{0}$ for $r > 1$, then:*

$$\max_{o \in \mathcal{O}} \left(1 + \overline{\Delta}\omega^o(r+1), \frac{1}{1 + \underline{\Delta}\omega^o(r+1)}\right) < \max_{o \in \mathcal{O}} \left(1 + \overline{\Delta}\omega^o(r), \frac{1}{1 + \underline{\Delta}\omega^o(r)}\right).$$

The above lemma suggests that when slices have not reached an equilibrium, then in the next round

$$\max_{o \in \mathcal{O}} \left(1 + \overline{\Delta}\omega^o(r+1), \frac{1}{1 + \underline{\Delta}\omega^o(r+1)}\right)$$

will decrease. This in turn suggests that maximum and minimum components of the vector of relative changes, $\mathbf{1} + \mathbf{\Delta \boldsymbol{\omega}}(r)$, are getting closer to 1.

With this result in hand, one can show the existence of a Lyapunov function guaranteeing convergence of the best response game, thus completing the proof of Theorem 3 – see Appendix for a detailed proof.

## IV. PERFORMANCE BOUNDS ANALYSIS

In this section we analyze the performance of the Nash equilibrium in terms of two standard metrics for efficiency and fairness: (*i*) the *price of anarchy*, which gives the loss in overall utility resulting from slices' strategic behavior, and (*ii*) *envy-freeness*, which captures the degree to which a slice would prefer another slice's allocations across the network resources. We will focus on the case where slice utilities are 1-fair homogeneous i.e., $U^o(\mathbf{w}) = \sum_{u \in \mathcal{U}^o} \phi_u \log(r_u(\mathbf{w})) \ \forall o \in \mathcal{O}$ – a widely accepted case leading to the well-known proportionally fair allocations.

### A. Efficiency: Price of Anarchy

The following result characterizes the socially optimal allocation of resources defined in Section II-C (see [22]).

**Fact 1.** *For slices with 1-fair homogenous utilities, the socially optimal allocation of resources $\mathbf{w}^*$ is such that $w_u^* = \phi_u s_o, \ \forall u \in \mathcal{U}^o$ and $\forall o \in \mathcal{O}$.*

The following theorem bounds the difference between the *overall network utility* resulting from the socially optimal allocation, $U(\mathbf{w}^*)$, and that obtained at a Nash equilibrium of the SCPA resource allocation mechanism, $U(\mathbf{w})$ – a sketch of the proof is provided in the Appendix.

**Theorem 4.** *If all slices have 1-fair homogenous utilities, then the Price of Anarchy (PoA) associated with a Nash equilibrium* **w** *satisfies*

$$PoA := U(\mathbf{w}^*) - U(\mathbf{w}) \leq \log(e).$$

*Furthermore, there exists a game instance for which this bound is tight.*

Note that, with 1-fair utilities, if we increase the capacity of all resources by a factor $\Delta c$, we have a utility increase of $\log(\Delta c)$. Thus, the performance improvement achieved by the socially optimal allocation over SCPA is (in the upper bound) equivalent to having a capacity $e$ times larger, i.e., almost the triple capacity. While there are some (pathological) cases in which such a bound can be achieved, our simulation results show that for practical scenarios the actual performance difference between the two allocations is much smaller, confirming that (for $\alpha_o = 1$) the flexibility gained with the SCPA mechanism comes at a very small price in performance.

*B. Fairness: Envy-freeness*

Next we consider a Nash equilibrium **w** and analyze whether a slice, say $o$, with utility $U^o(\mathbf{w})$, might have a better utility if it were to exchange its resources with those of another slice, say $o'$. To that end, we denote by $\tilde{\mathbf{w}}$ the resulting weight allocation when the users of slices $o$ and $o'$ exchange their allocated resources. It is easy to see that $\tilde{\mathbf{w}}^o$ is such that

$$\tilde{w}_u^o = \frac{\phi_u}{\sum_{v \in \mathcal{U}_b^o} \phi_v} d_b^{o'}(\mathbf{w}) \text{ for all } b \in \mathcal{B} \text{ and all } u \in \mathcal{U}_b^o, \quad (3)$$

i.e., slice $o$ takes the aggregate weight of $o'$ at base station $b$ under the Nash equilibrium, $d_b^{o'}(\mathbf{w})$, and allocates it proportionally to its user priorities. Clearly, $\tilde{\mathbf{w}}^{o'}$ is defined similarly and the remaining slices weights remain unchanged under $\tilde{\mathbf{w}}$.

We define the envy of slice $o$ for $o'$'s resources under the Nash equilibrium **w** by

$$E^{o,o'} := U^o(\tilde{\mathbf{w}}) - U^o(\mathbf{w}).$$

Note that envy is a "directed" concept, i.e., it is defined from slice $o$'s point of view. When $E^{o,o'} \leq 0$, we say slice $o$ is not envious. The following theorem provides a bound on $E^{o,o'}$ – see the Appendix for a sketch of the proof.

**Theorem 5.** *Consider a slice $o$ with 1-fair homogeneous utilities and the remaining slices $\mathcal{O} \setminus \{o\}$ with arbitrary slice utilities. Consider a slice $o'$ such that $s_o = s_{o'}$. Let* **w** *denote a Nash equilibrium and $\tilde{\mathbf{w}}$ denote the resulting weights when $o$ and $o'$ exchange their resources. Then, the envy of slice $o$ for $o'$ satisfies*

$$E^{o,o'} = U^o(\tilde{\mathbf{w}}) - U^o(\mathbf{w}) \leq 0.060.$$

*Furthermore, there is a game instance where $0.041 \leq E_{o,o'}$.*

Given that, if one increases the rates of all users by a factor $\Delta r$ this yields a utility increase of $\log(\Delta r)$, one can interpret this result as saying that, by exchanging resources with $o'$, slice $o$ may see a gain equivalent to increasing the rate of all its users by a factor between $4.1\%$ and $6.1\%$ (given by the lower and upper bounds of the above theorem). This is quite low and, moreover, simulation results show that in practical settings there is actually (almost) never any envy, confirming that our system is (practically) envy-free.

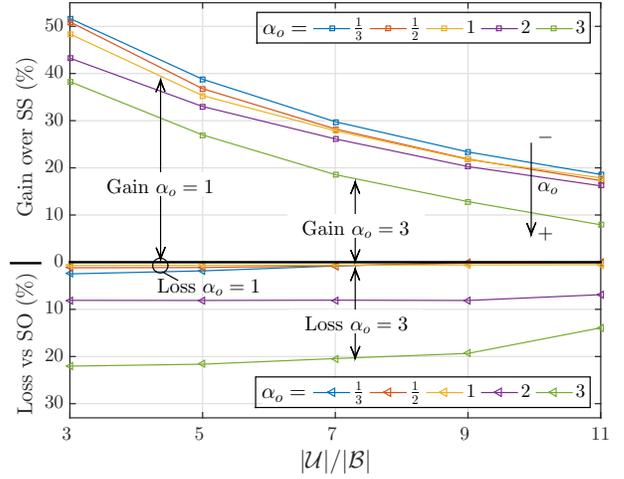

Fig. 1: Average Gain over Static slicing and Loss against Social optimum for different scenarios.

## V. PERFORMANCE EVALUATION

Next, we evaluate the performance of the SCPA resource allocation mechanism via simulation. The mobile network scenario considered is based on the IMT-A evaluation guidelines for dense 'small cell' deployments [23], which consider base stations with an intersite distance of 200 meters in a hexagonal cell layout with 3 sector antennas.[5] The network size $|\mathcal{B}|$ is 57 sectors and, unless otherwise stated, users move according to the Random Waypoint Model (RWP). Users' Signal Interference to Noise Ratio ($\overline{\text{SINR}}_u$) is computed based on physical layer network model specified in [23] (which includes path loss, shadowing, fast fading and antenna gain) and user association follows the strongest signal policy. The achievable rate for users, $c_u$, are determined based on the thresholds reported in [24]. For all our simulation results, we obtained $95\%$ confidence intervals with relative errors below $1\%$ (not shown in the figures).

*A. Overall performance*

Throughout the paper we have used *static slicing* and the *socially optimal* resource allocations as our baselines. In order to confirm our analytical results and gain additional insights, we have evaluated the performance of the SCPA mechanism versus these two baselines via simulation. As an intuitive metric for comparison, we have used the extra capacity required by these baseline schemes to deliver the same performance as SCPA:

($i$) *Gain over SS*: additional resources required by *static slicing* to provide the same utility as SCPA (in %).

---

[5]Note that, in this setting, users associate with sectors rather than the base stations we used in the mechanism description and analysis.

(ii) *Loss versus SO*: additional resources required by SCPA to provide the same utility as the *socially optimal* allocation (in %); note that this metric is closely related to the *Price of Anarchy* analyzed in Section IV-A.

The results shown in Figure 1 are for different user densities ($|\mathcal{U}|/|\mathcal{B}|$) and different slice utilities ($\alpha_o$ parameter). As expected, the SCPA mechanism always has a gain over static slicing and a loss over the social optimal. However, for $\alpha_o = 1$ the loss is well below the bound given in Section IV-A. We further observe that performance is particularly good as long as $\alpha_o$ does not exceed 1 (Gain over SS up to 50% and Loss over SO below 5%), and it degrades mildly as $\alpha_o$ increases.

### B. Fairness

In addition to overall performance, it is of interest to evaluate the fairness of the resulting allocations. While in Section IV-B we derived analytically a bound on the envy, we have further explored this via simulation by evaluating up to $10^7$ randomly generated scenarios, with parameters drawn uniformly in the ranges: $|\mathcal{O}| \in [2, 12]$, $|\mathcal{B}| \in [10, 90]$, $|\mathcal{U}|/|\mathcal{B}| \in [3, 15]$, $\alpha_o \in [0.01, 30]$ and $\phi$ vectors in the simplex. Our results show that $E^{o,o'} < 0$ holds for *all* the cases explored, confirming that in practice the system is envy-free.

### C. Protection against other slices

One of the main objectives of our proposed framework is to enable slices to customize their resource allocations. This can be done by adjusting (i) the user priorities $\phi_u$, and (ii) the parameter $\alpha_o$, which regulates the desired level of fairness among the slice's users. In order to evaluate the impact that these settings have amongst slices, we simulated a scenario with three slices: Slice 1 has $\alpha_1 = 1$, Slice 2 has $\alpha_2 = 4$, and Slice 3 has $\alpha_3$ with varying values. For simplicity, we set the priorities $\phi_u$ equal for all users.

Figure 2 shows the rate distributions of the 3 slices. We observe that the choice of $\alpha_3$ is effective in adjusting the level of user fairness for Slice 3; indeed, as $\alpha_3$ grows, the rate distribution becomes more homogeneous. Such customization at Slice 3 has a higher impact on Slice 1 than on Slice 2. This is the case because, as $\alpha_2$ is quite large, the distribution of Slice 2's rates remains homogeneous, making the slice fairly insensitive to the choices of the other slices. As can be seen in the subplots, the utilities of Slices 1 and 2 are not only larger than the utility of static slicing, but remain fairly insensitive to $\alpha_3$, showing that in both cases we have a good level of protection between slices.

### D. Convergence speed

The existence of a Nash equilibrium and the convergence of Best Response Dynamics are essential for the system stability. While the existence of a Nash equilibrium has been proved for all $\alpha_o > 0$, convergence has only been shown for $\alpha_o \in [1, 2]$. In order to confirm the convergence of best responses for other $\alpha_o$ values, we have conducted extensive simulations involving up to $10^7$ randomly generated scenarios in the same parameter space as in Section V-B. Our results confirmed the convergence of the best response game in all cases. Moreover, they also showed that convergence speed mainly depends on $\alpha_o$, while it is fairly insensitive to the user priorities and the network size. According to the results, convergence is very quick for $\alpha_o \leq 1$ (about 8 rounds) and increases slightly as $\alpha_o$ grows (about 16 rounds for $\alpha_o = 3$). The average number of rounds needed for the Best Response dynamics to converge are shown in Figure 3.

### E. Impact of user mobility

The above performance results assumes a Random Waypoint model in which, on average, users distributions are uniform. In order to understand the impact of other users distributions, we evaluated the *Gain over SS* for scenarios with 4 slices, equal shares and four different users mobility patterns:

1) *Uniform*: homogeneous slices with uniform spatial loads.
2) *Overlapping hotspots*: homogeneous slices with the same non-uniform spatial loads.
3) *Non-overlapping hotspots*: heterogenous slices with orthogonal non-uniform spatial loads which when aggregate are more uniform.
4) *Mixed*: two heterogenous slices with the same non-uniform spatial loads, two slices with uniform spatial loads.

Results of our simulations depicted in Figure 4 indicate that *Gains over SS* are bigger at scenarios with uneven and complementary traffic loads, since the different slices are capable to obtain dynamically a resource allocation that better adapts to the slices traffic distribution. Moreover, we can see that a higher $\alpha$ reports a smaller gain over Static Slicing, since the slices tend to be more fair amongst its users and the ability to exploit statistical multiplexing gets reduced.

## VI. CONCLUSIONS

In this paper we have analyzed a 'share-constrained proportional allocation' framework for network slicing. The framework allows slices to customize the resource allocation to their users, leading to a *network slicing game* in which each slice reacts to the settings of the others. Our main conclusion is that the framework provides an *effective* and *implementable* scheme for dynamically sharing resources across slices. Indeed, this scheme involves simple operations at base stations and incurs a limited signaling between the slices and the infrastructure. Our results confirm system stability (*best response dynamics* converge), substantial gains over *static slicing*, and fairness of the allocations (*envy-freeness*). Moreover, as long as the majority of the slices do not choose $\alpha_o$ values larger than 1 (i.e., they do not all demand very homogeneous rate distributions), the overall performance is close to optimal (*price of anarchy* is very small). Thus, in this case the flexibility provided by this framework comes at no cost. If a substantial number of slices choose higher $\alpha_o$'s, then we pay a (small) price for enabling slice customization.

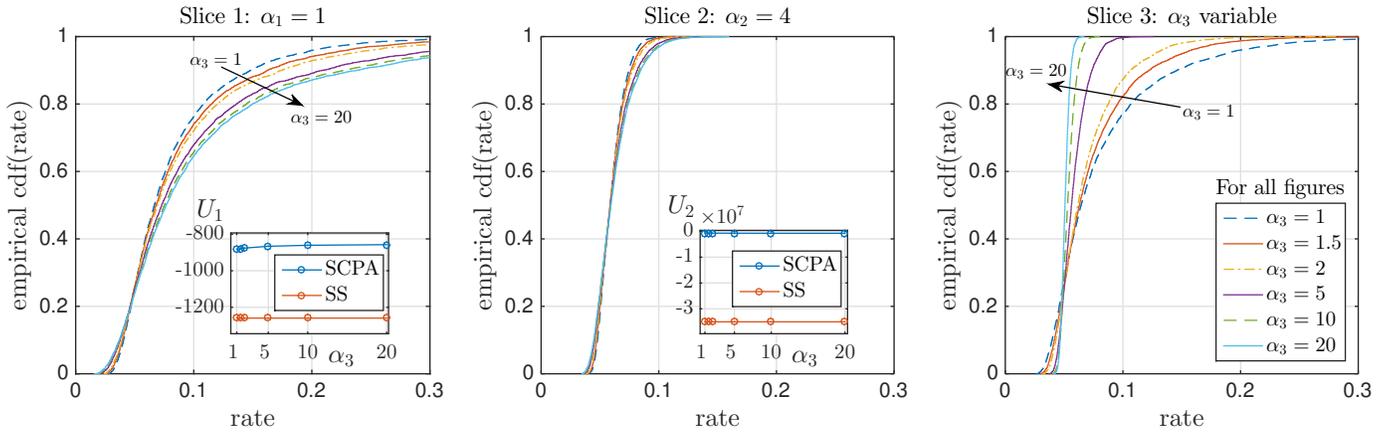

Fig. 2: Impact of $\alpha_3$ decision on the slice rate distributions.

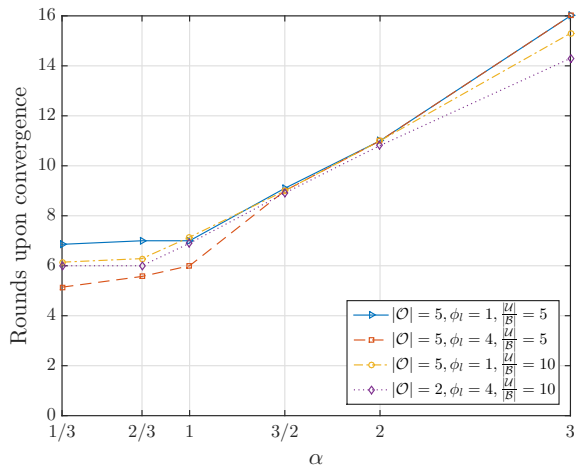

Fig. 3: Average number of round upon convergence for different scenarios.

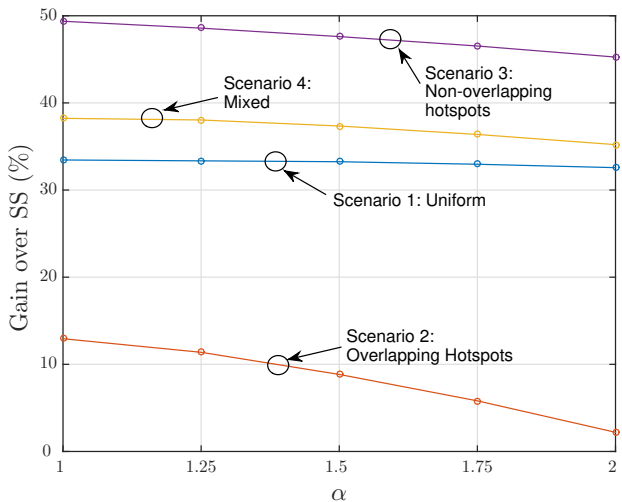

Fig. 4: Gain over Static slicing for different traffic models and alpha values.

ACKNOWLEDGEMENTS

The work of P. Caballero and G. de Veciana has been supported by NSF Award CNS1232283, and the work of A. Banchs and X. Costa-Pérez by the H2020-ICT-2014-2 project 5G NORMA under Grant Agreement No. 671584.

## APPENDIX

### Proof of Lemma 1

Given the weight allocation under static slicing, $\mathbf{w}^{ss}$, and the weights of the other slices under dynamic sharing, $\mathbf{w}^{-o}$, we consider the following weight allocation for slice $o$:

$$w_u = \frac{w_u^{ss}}{\sum_{u\in\mathcal{U}_{b(u)}^o} w_u^{ss}} \frac{a_{b(u)}^o(\mathbf{w}^{-o})}{\sum_{b'\in\mathcal{B}_o} a_{b'}^o(\mathbf{w}^{-o})} s_o, \quad (4)$$

where $\mathcal{B}_o$ is the set of base stations where slice $o$ has customers.

We define $\rho_u^o(\mathbf{w}^{ss})$ as the ratio between the weight of user $u$ under static slicing and the sum of the weights of all the users of the same slice in the base station, i.e.,

$$\rho_u^o(\mathbf{w}^{ss}) \doteq \frac{w_u^{ss}}{\sum_{u\in\mathcal{U}_{b(u)}^o} w_u^{ss}}$$

where we have dropped the terms $\mathbf{w}^{-o}$ and $\mathbf{w}^{ss}$ from $a_{b(u)}^o(\mathbf{w}^{-o})$ and $\rho_u^o(\mathbf{w}^{ss})$ for readability purposes.

With the allocation given by (4), for two users $u$ and $u'$ of slice $o$ it holds

$$\frac{w_u}{w_{u'}} = \frac{\rho_u^o}{\rho_{u'}^o} \frac{a_{b(u)}^o}{a_{b(u')}^o} \quad (5)$$

Furthermore, it also holds that

$$d_b^o = \sum_{u\in\mathcal{U}_b^o} w_u = \sum_{u\in\mathcal{U}_b^o} \rho_u^o \frac{a_{b(u)}^o}{\sum_{b'\in\mathcal{B}_o} a_{b'}^o} s_o = \frac{a_{b(u)}^o}{\sum_{b'\in\mathcal{B}_o} a_{b'}^o} s_o = \frac{w_u}{\rho_u^o}$$

for $u \in \mathcal{U}_b^o$.

From the above expression, we have

$$\frac{\rho_u^o l_{b(u)}(\mathbf{w})}{\rho_{u'}^o l_{b(u')}(\mathbf{w})} = \frac{\rho_u^o\left(a_{b(u)}^o + d_{b(u)}^o\right)}{\rho_{u'}^o\left(a_{b(u')}^o + d_{b(u')}^o\right)} = \frac{\rho_u^o\left(a_{b(u)}^o + \frac{w_u}{\rho_u^o}\right)}{\rho_{u'}^o\left(a_{b(u')}^o + \frac{w_{u'}}{\rho_{u'}^o}\right)},$$

and combining this with (5):

$$\frac{\rho_u^o l_{b(u)}(\mathbf{w})}{\rho_{u'}^o l_{b(u')}(\mathbf{w})} = \frac{\rho_u^o\left(a_{b(u)}^o + \frac{a_{b(u)}^o}{a_{b(u')}^o}\frac{w_{u'}}{\rho_{u'}^o}\right)}{\rho_{u'}^o\left(a_{b(u')}^o + \frac{w_{u'}}{\rho_{u'}^o}\right)}$$

$$= \frac{\rho_u^o a_{b(u)}^o \left(1 + \frac{w_{u'}}{a_{b(u')}^o \rho_{u'}^o}\right)}{\rho_{u'}^o a_{b(u')}^o \left(1 + \frac{w_{u'}}{a_{b(u')}^o \rho_{u'}^o}\right)} = \frac{\rho_u^o a_{b(u)}^o}{\rho_{u'}^o a_{b(u')}^o}$$

From the above,

$$w_u = \frac{w_u}{\sum_{u'\in\mathcal{U}^o} w_{u'}} s_o = \frac{s_o}{\sum_{u'\in\mathcal{U}^o} \frac{w_{u'}}{w_u}} =$$
$$= \frac{s_o}{\sum_{u'\in\mathcal{U}^o} \frac{\rho_{u'}^o a_{b(u')}^o}{\rho_u^o a_{b(u)}^o}} = \frac{s_o}{\sum_{u'\in\mathcal{U}^o} \frac{\rho_{u'}^o l_{b(u')}(\mathbf{w})}{\rho_u^o l_{b(u)}(\mathbf{w})}}$$
$$= \frac{\rho_u^o l_{b(u)}(\mathbf{w})}{\sum_{u'\in\mathcal{U}^o} \rho_{u'}^o l_{b(u')}(\mathbf{w})} s_o = \frac{\rho_u^o l_{b(u)}(\mathbf{w})}{\sum_{b'\in\mathcal{B}_o} l_{b'}(\mathbf{w})} s_o$$

Since $\mathcal{B}_o \subseteq \mathcal{B}$:

$$\sum_{b\in\mathcal{B}_o} l_b(\mathbf{w}) \leq \sum_{b\in\mathcal{B}} l_b(\mathbf{w}) = 1$$

and thus

$$w_u \geq \rho_u^o l_{b(u)}(\mathbf{w}) s_o,$$

from which

$$r_u(\mathbf{w}) = \frac{w_u}{l_{b(u)}(\mathbf{w})} c_u \geq \frac{\rho_u^o l_{b(u)}(\mathbf{w}) s_o}{l_{b(u)}(\mathbf{w})} c_u = \rho_u^o s_o c_u = r_u^{ss}.$$

The above holds for all $u \in \mathcal{U}$, which proves the lemma.

### Proof of Theorem 1

This result follows from Lemma 1. Given the configuration of the other slices, there exists a configuration for a given slice under which all its users obtain at least the same throughput as with static slicing, and thus the slice's utility with this configuration is at least as high. As a consequence, in a Nash equilibrium the slice will receive a utility no smaller than this value.

### Proof of Lemma 2

Let us start for $\alpha_o \neq 1$. The best response of slice $o$ is given by

$$\mathbf{w}^o = \arg\max_{\mathbf{w}'^o} \sum_{u\in\mathcal{U}_o} \frac{\phi_u}{1-\alpha_o}\left(\frac{w_u' c_u}{a_{b(u)}^o(\mathbf{w}'^{-o}) + d_{b(u)}^o(\mathbf{w}'^o)}\right)^{1-\alpha_o}$$

$$\text{subject to: } \sum_{u\in\mathcal{U}_o} w_u' = s^o, \quad w_u' \geq 0, \quad \forall u \in \mathcal{U}_o$$

The Lagrangian for this optimization problem is given by

$$\mathcal{L}(\mathbf{w}, \lambda) = \sum_{u\in\mathcal{U}_o} \frac{\phi_u}{1-\alpha_o}\left(\frac{w_u c_u}{a_b^o(\mathbf{w}^{-o}) + d_b^o(\mathbf{w}'^o)}\right)^{1-\alpha_o}$$
$$- \lambda_o\left(\sum_{u\in\mathcal{U}_o} w_u - s_o\right)$$

The partial derivative of the above function with respect to $w_u$ is given by

$$\frac{\partial \mathcal{L}(\mathbf{w}, \lambda)}{\partial w_u} = \frac{\phi_u c_u^{1-\alpha_o} \cdot (l_b(\mathbf{w}) - w_u)}{w_u^{\alpha_o} \cdot l_b(\mathbf{w})^{2-\alpha_o}}$$

$$-\sum_{u' \in \mathcal{U}_{b(u)}^o \setminus \{u\}} \frac{\phi_{u'} c_{u'}^{1-\alpha_o} w_{u'}^{1-\alpha_o}}{l_b(\mathbf{w})^{2-\alpha_o}} - \lambda_o$$

$$= \frac{\phi_u c_u^{1-\alpha_o} \cdot l_b(\mathbf{w})}{w_u^{\alpha_o} \cdot l_b(\mathbf{w})^{2-\alpha_o}} - \sum_{u' \in \mathcal{U}_{b(u)}^o} \frac{\phi_{u'} c_{u'}^{1-\alpha_o} w_{u'}^{1-\alpha_o}}{l_b(\mathbf{w})^{2-\alpha_o}} - \lambda_o$$

From the above, for two users $u$, $u'$ in the same base station we have

$$w_u^{\alpha_o} = \frac{\phi_u}{\phi_{u'}} \left( \frac{c_u}{c_{u'}} \right)^{1-\alpha_o} w_{u'}^{\alpha_o} \quad (6)$$

Combining the above two equations yields

$$\frac{\partial \mathcal{L}(\mathbf{w}, \lambda)}{\partial w_u} = \frac{\phi_u \cdot c_u^{1-\alpha_o}}{w_u^{\alpha_o} \cdot l_b(\mathbf{w})^{2-\alpha_o}} \times$$

$$\left( l_b(\mathbf{w}) - \sum_{u' \in \mathcal{U}_{b(u)}^o} \frac{\phi_{u'} c_{u'}^{1-\alpha_o}}{\phi_u c_u^{1-\alpha_o}} w_{u'}^{1-\alpha_o} w_u^{\alpha_o} \right) - \lambda_o$$

$$= \frac{\phi_u \cdot c_u^{1-\alpha_o}}{w_u^{\alpha_o} \cdot l_b(\mathbf{w})^{2-\alpha_o}} \left( l_b(\mathbf{w}) - \sum_{u' \in \mathcal{U}_{b(u)}^o} w_{u'} \right) - \lambda_o.$$

Equaling the above expression for two different users $u$ and $u'$ at different base stations, we obtain the following expression, which holds for any pair of users of slice $o$ (at the same or different base stations):

$$\frac{w_u}{w_{u'}} = \frac{\beta_u}{\beta_v} \frac{\frac{(a_{b(u)}^o(\mathbf{w}^{-o}))^{\frac{1}{\alpha_o}}}{(a_{b(u)}^o(\mathbf{w}^{-o}) + d_{b(u)}^o(\mathbf{w}^o))^{\frac{2}{\alpha_o}-1}}}{\frac{(a_{b(u')}^o(\mathbf{w}^{-o}))^{\frac{1}{\alpha_o}}}{(a_{b(u')}^o(\mathbf{w}^{-o}) + d_{b(u')}^o(\mathbf{w}^o))^{\frac{2}{\alpha_o}-1}}} \quad (7)$$

where $\beta_u := (\phi_u)^{\frac{1}{\alpha_o}} (c_u)^{\frac{1}{\alpha_o}-1}$. From the above, we obtain (2) by normalizing. In order to prove that the resulting non-linear system of equations has a unique solution, we proceed as follows. Let $d_b^o = \sum_{u \in \mathcal{U}_b^o} w_u$. From (6),

$$d_b^o = w_u \sum_{u' \in \mathcal{U}_b^o} \frac{\beta_{u'}}{\beta_u}$$

Combining the above with (7) yields

$$\frac{d_b^o}{d_{b'}^o} = \frac{\beta_u}{\beta_v} \frac{\sum_{v' \in \mathcal{U}_b^o} \frac{\beta_{v'}}{\beta_v} \frac{(a_b^o(\mathbf{w}^{-o}))^{\frac{1}{\alpha_o}}}{(a_b^o(\mathbf{w}^{-o}) + d_b^o)^{\frac{2}{\alpha_o}-1}}}{\sum_{v' \in \mathcal{U}_{b'}^o} \frac{\beta_{v'}}{\beta_u} \frac{(a_{b'}^o(\mathbf{w}^{-o}))^{\frac{1}{\alpha_o}}}{(a_{b'}^o(\mathbf{w}^{-o}) + d_{b'}^o)^{\frac{2}{\alpha_o}-1}}}$$

which is equivalent to

$$d_b^o \left( a_b^o(\mathbf{w}^{-o}) + d_b^o \right)^{\frac{2}{\alpha_o}-1} = K d_{b'}^o \left( a_{b'}^o(\mathbf{w}^{-o}) + d_{b'}^o \right)^{\frac{2}{\alpha_o}-1}$$

where $K$ is some constant. Then, if we fix $d_b^o$ to some positive value, there exists a unique positive value of $d_{b'}^o$ that satisfies the above equation. Indeed, the lhs of the equation will be fixed to some finite value larger than 0, while the rhs grows from 0 to $\infty$ as we increase $d_{b'}^o$. Moreover, the larger the value of $d_b^o$, the larger the resulting $d_{b'}^o$, since both the lhs and the rhs of the equation are increasing functions of $d_b^o$ and $d_{b'}^o$, respectively.

From the above, we can compute the $d_{b'}^o$ value of each base stations as a function of a single $d_b^o$. Once we have all $d_{b'}^o$ values, we can uniquely compute the user weights $w_u$, which are an increasing function of $d_{b'}^o$ (and thus of $d_b^o$). Inserting the resulting weights into $\sum_{u \in \mathcal{U}_o} w_u(d_b^o) = s_o$, we have an equation with a single unknown, $d_b^o$. This equation has a unique solution, as the lhs is an increasing function of $d_b^o$ and the rhs is constant. Computing the resulting $d_b^o$ value, and obtaining from this value the corresponding $w_u$ values, we have a solution to the system. Since all relationships are bijective, this is the only solution of the system.

The case $\alpha_o = 1$ is proven employing a similar argument. Indeed, by repeating the same steps as above for for $U_o(\mathbf{w}) = \sum_{u \in \mathcal{U}_o} \phi_u \log r_u(\mathbf{w})$, it is easy to verify the best response for this case corresponds to the expression given by (2) for $\alpha_o = 1$.

*Proof of Theorem 2*

We start by proving the following lemma which is required for the proof of the theorem.

**Lemma 4.** *Let $\mathbf{R}_o$ be the implicit function that denotes the best response of slice $o$, i.e., $\mathbf{w}^o = \mathbf{R}_o(\mathbf{w}^o, \mathbf{w}^{-o})$. $\mathbf{R}^o$ is a continuous and differentiable function for $a_b^o \neq 0 \; \forall b$.*

*Proof.* We start by noting that $\mathbf{R}_o$ is a continuous and differentiable function of $\boldsymbol{a}^o = (a_b^o : b \in \mathcal{B}_o)$ when $\boldsymbol{w}^o$ is fixed. So it follows from the implicit function theorem that the best response is continuous and differentiable in $\boldsymbol{a}^o$. As $\boldsymbol{a}^o$ is continuous function of $\mathbf{w}^{-o}$, it follows that the best response is a continuous and differentiable function of the other slices' weights. □

Now, let us define a perturbed game $G^{(\varepsilon)}$ for some $\varepsilon > 0$, in which there is an additional slice which places a weight of $\varepsilon$ in each base station.[6] The existence of a NE for this perturbed game is guaranteed by the result of [21] since $(i)$ the utility is a concave function, $(ii)$ it is continuous (as given by the previous lemma),[7] and $(iii)$ the game strategy space set of a slice, given by $\mathcal{S}^o = \{\mathbf{w}^o \in \mathbf{R}^o \mid \mathbf{w}^o \geq \mathbf{0} \text{ and } \sum_{u \in \mathcal{U}^o} w_u = s^o\}$, is compact and convex.[8] Moreover, the following lemma shows that the weights in the NE must be greater than or equal to a value $\delta$ for any $\varepsilon < \varepsilon_{max}$.[9]

**Lemma 5.** *For any NE weight allocation of game $G^{(\varepsilon)}$, there exists a constant $\delta > 0$ such that $w_u^\varepsilon \geq \delta \; \forall u \in \mathcal{U}$ and $\varepsilon < \varepsilon_{max}$.*

---

[6] In the perturbed game, the shares are rescaled so that they still sum 1, i.e., $\sum_{o \in \mathcal{O}} s_o + |\mathcal{B}|\varepsilon = 1$.
[7] Note that in the perturbed game $a_b^o \geq \varepsilon > 0$.
[8] The statement of the theorem of [21] requires that $U_o(\mathbf{w})$ is defined in the strategy space $\mathcal{S}^o$, which is not satisfied for $\alpha_o = 1$ since in this case $U_o(\mathbf{w}) \to -\infty$ when $w_u = 0$ for some $u \in \mathcal{U}_o$. However, the proof of this theorem only requires that the mapping $\mathbf{w}^o = R_o(\mathbf{w}^{-o})$ is defined in $\mathcal{S}^o$, which is satisfied for $\alpha_o = 1$ (and indeed this mapping never yields $w_u = 0$, as this would not maximize the slice's utility).
[9] Note that, as the sum of the weights of all players in the game is 1, i.e., $\sum_{o \in \mathcal{O}} s_o + |\mathcal{B}|\varepsilon = 1$, we have the following upper bound for $\varepsilon$: $\varepsilon \leq \hat{\varepsilon} = 1/|\mathcal{B}|$.

*Proof.* According to Lemma 2, the best response of a user $u \in \mathcal{U}^o$ in the $\varepsilon$ perturbed game is given by

$$w_u = \frac{\frac{(\varepsilon+a_{b(u)}^o)^{\frac{1}{\alpha_o}}}{(\varepsilon+a_{b(u)}^o+d_{b(u)}^o)^{\frac{2}{\alpha_o}-1}}}{\sum_{v \in \mathcal{U}^o} \left(\frac{\beta_v}{\beta_u}\right) \frac{(\varepsilon+a_{b(v)}^o)^{\frac{1}{\alpha_o}}}{(\varepsilon+a_{b(v)}^o+d_{b(v)}^o)^{\frac{2}{\alpha_o}-1}}} s_o. \quad (8)$$

In order to derive a bound for $w_u$, we proceed along the following steps. First, we obtain a bound for $w_u$ as a function of $l_b$. Second, we derive a bound for $l_b$. Finally, by combining the results of the first and the second steps, we obtain a bound for $w_u$.

*Bound for $w_u$ as a function of $l_b$*

We will first prove the existence of a bound for the case of $\alpha_o \geq 1$ and then for the case $\alpha_o < 1$. Let us start with the case $\alpha_o \geq 1$. From (2) it follows

$$w_u \geq \frac{\frac{(\varepsilon+a_{b(u)}^o)^{\frac{1}{\alpha_o}}}{(\varepsilon+a_{b(u)}^o+d_{b(u)}^o)^{\frac{2}{\alpha_o}-1}}}{\max_{v \in \mathcal{U}^o}\left[\frac{\beta_v}{\beta_u}\right] \sum_{v \in \mathcal{U}^o} \frac{(\varepsilon+a_{b(v)}^o)^{\frac{1}{\alpha_o}}}{(\varepsilon+a_{b(v)}^o+d_{b(v)}^o)^{\frac{2}{\alpha_o}-1}}} s_o$$

Let us define the constant $m_u^o = \max_{v \in \mathcal{U}^o}\left[\frac{\beta_v}{\beta_u}\right]$. Then,

$$w_u \geq \frac{\frac{(\varepsilon+l_{b(u)}-d_{b(u)}^o)^{\frac{1}{\alpha_o}}}{(\varepsilon+l_{b(u)})^{\frac{2}{\alpha_o}-1}}}{m_u^o \sum_{v \in \mathcal{U}^o} \frac{(\varepsilon+l_{b(v)}-d_{b(v)}^o)^{\frac{1}{\alpha_o}}}{(\varepsilon+l_{b(v)})^{\frac{2}{\alpha_o}-1}}} s_o$$

$$\geq \frac{\frac{(l_{b(u)}-d_{b(u)}^o)^{\frac{1}{\alpha_o}}}{(l_{b(u)})^{\frac{2}{\alpha_o}-1}}}{m_u^o \sum_{v \in \mathcal{U}^o} (\varepsilon+l_{b(v)})^{-\frac{1}{\alpha_o}+1}} s_o$$

$$\geq \frac{\frac{(l_{b(u)}-d_{b(u)}^o)^{\frac{1}{\alpha_o}}}{(l_{b(u)})^{\frac{2}{\alpha_o}-1}}}{m_u^o \sum_{v \in \mathcal{U}^o} \left((\varepsilon)^{-\frac{1}{\alpha_o}+1} + (l_{b(v)})^{-\frac{1}{\alpha_o}+1}\right)} s_o$$

$$\geq \frac{\frac{(l_{b(u)}-d_{b(u)}^o)^{\frac{1}{\alpha_o}}}{(l_{b(u)})^{\frac{2}{\alpha_o}-1}}}{m_u^o \sum_{v \in \mathcal{U}^o} \left(1 + (\varepsilon)^{-\frac{1}{\alpha_o}+1}\right)} s_o$$

$$\geq \frac{s_o}{2m_u^o|\mathcal{U}^o|} \frac{(l_{b(u)}-d_{b(u)}^o)^{\frac{1}{\alpha_o}}}{(l_{b(u)})^{\frac{2}{\alpha_o}-1}}$$

We next focus on the case $\alpha_o < 1$. From (2), it follows that

$$\frac{(w_u)^{\alpha_o}}{(w_v)^{\alpha_o}} = \frac{\frac{(\varepsilon+l_{b(u)}-d_{b(u)}^o)}{(\varepsilon+l_{b(u)})^{2-\alpha_o}}}{\frac{(\varepsilon+l_{b(v)}-d_{b(v)}^o)}{(\varepsilon+l_{b(v)})^{2-\alpha_o}}} \geq \frac{\frac{(l_{b(u)}-d_{b(u)}^o)}{(l_{b(u)})^{2-\alpha_o}}}{\frac{(\varepsilon+l_{b(v)}-d_{b(v)}^o)}{(\varepsilon+l_{b(v)})^{2-\alpha_o}}}$$

From the above,

$$(w_u)^{\alpha_o} \geq \frac{(l_{b(u)}-d_{b(u)}^o)}{(l_{b(u)})^{2-\alpha_o}} \frac{(\varepsilon+l_{b(v)})^{2-\alpha_o}}{(\varepsilon+l_{b(v)}-d_{b(v)}^o)}(w_v)^{\alpha_o}$$

$$\geq \frac{(l_{b(u)}-d_{b(u)}^o)}{(l_{b(u)})^{2-\alpha_o}}(\varepsilon+l_{b(v)})^{1-\alpha_o}(w_v)^{\alpha_o}$$

$$\geq \frac{(l_{b(u)}-d_{b(u)}^o)}{(l_{b(u)})^{2-\alpha_o}} w_v \left(\frac{w_v}{l_{b(u)}}\right)^{1-\alpha_o}$$

$$\geq \frac{(l_{b(u)}-d_{b(u)}^o)}{(l_{b(u)})^{2-\alpha_o}} w_v$$

Given that the sum of the weights of all the users of slice $o$ is equal to $s_o$, there needs to be at least one user $v$ for which $\sum_{u' \in \mathcal{U}_{b(v)}^o} w_{u'} = d_v w_v \geq \frac{s_o}{|\mathcal{U}^o|}$. If we take such a user $v$, from the above we obtain

$$(w_u)^{\alpha_o} \geq \frac{(l_{b(u)}-d_{b(u)}^o)}{(l_{b(u)})^{2-\alpha_o}} \frac{s_o}{|\mathcal{U}^o|}, \quad \forall u \in \mathcal{U}_o$$

Isolating $w_u$ from the above yields

$$w_u \geq \left(\frac{s_o}{|\mathcal{U}_o|m_u^o}\right)^{\frac{1}{\alpha_o}} \frac{(l_{b(u)}-d_{b(u)}^o)^{\frac{1}{\alpha_o}}}{(l_{b(u)})^{\frac{2}{\alpha_o}-1}}$$

$$\geq \frac{s_o}{2m_u^o|\mathcal{U}^o|} \frac{(l_{b(u)}-d_u w_u)^{\frac{1}{\alpha_o}}}{(l_{b(u)})^{\frac{2}{\alpha_o}-1}}$$

Putting together the bounds obtained for $\alpha_o \geq 1$ and $\alpha_o < 1$ leads to the following expression which holds for any $\alpha_o > 0$:

$$w_u \geq \frac{s_o}{2m_u^o|\mathcal{U}^o|} \frac{(l_{b(u)}-d_u w_u)^{\frac{1}{\alpha_o}}}{(l_{b(u)})^{\frac{2}{\alpha_o}-1}}, \text{ for } \alpha_o > 0 \quad (9)$$

*Bound for $l_b$*

We will now provide a bound for $l_b$. Let us define $\mathcal{U}_b^*$ as a subset of $\mathcal{U}$ that contains a representative user of each slice at base station $b$. Let us further define sets $\mathcal{U}_b^{(*,\geq 1)}$ and $\mathcal{U}_b^{(*,<1)}$ as the subsets of $\mathcal{U}_b^*$ corresponding to the slices with $\alpha_o \geq 1$ and $\alpha_o < 1$, respectively. Note that

$$l_{b(u)} = \sum_{u \in \mathcal{U}_{b(u)}^{(*,\geq 1)}} d_u w_u = \sum_{u \in \mathcal{U}_{b(u)}^{(*,\geq 1)}} d_u w_u + \sum_{u \in \mathcal{U}_{b(u)}^{(*,<1)}} d_u w_u.$$

Let us denote the largest $\alpha_o$ of the slices with $\alpha_o \geq 1$ by $\overline{\alpha_o}$ and the smallest $\alpha_o$ of these slices by $\underline{\alpha_o}$. Similarly, let $\overline{\alpha_o'}$ denote the largest $\alpha_o$ of the slices with $\alpha_o < 1$ and $\underline{\alpha_o'}$ the smallest. Let us further define $\eta = \min_{o \in \mathcal{O}, u \in \mathcal{U}_{b(u)}^{(*,\geq 1)}} \frac{s_o d_u}{2m_u^o |\mathcal{U}|}$. Then, using the bound for $w_u$ in (9), we obtain

$$\sum_{u \in \mathcal{U}_{b(u)}^{(*,\geq 1)}} d_u w_u \geq \eta \sum_{u \in \mathcal{U}_{b(u)}^{(*,\geq 1)}} \frac{(l_{b(u)}-d_u w_u)^{\frac{1}{\alpha_o(u)}}}{(l_{b(u)})^{\frac{2}{\alpha_o(u)}-1}}$$

$$\geq \eta(l_{b(u)})^{1-\frac{1}{\alpha_o}} \sum_{u \in \mathcal{U}_{b(u)}^{(*,\geq 1)}} \left(1 - \frac{d_u w_u}{l_{b(u)}}\right)^{\frac{1}{\alpha_o}}$$

$$\geq \eta(l_{b(u)})^{1-\frac{1}{\alpha_o}} \left[\sum_{u \in \mathcal{U}_{b(u)}^{(*,\geq 1)}} 1 - \sum_{u \in \mathcal{U}_{b(u)}^{(*,\geq 1)}} \frac{d_u w_u}{l_{b(u)}}\right]^{\frac{1}{\alpha_o}}$$

$$\geq \eta(l_{b(u)})^{1-\frac{1}{\alpha_o}} \left[\sum_{u \in \mathcal{U}_{b(u)}^{(*,\geq 1)}} 1 - \sum_{u \in \mathcal{U}_{b(u)}^*} \frac{d_u w_u}{l_{b(u)}}\right]^{\frac{1}{\alpha_o}}$$

$$\geq \eta ( \sum_{u \in \mathcal{U}_{b(u)}^{(*,\geq 1)}} d_u w_u)^{1-\frac{1}{\alpha_o}} \left[ (|\mathcal{U}_{b(u)}^{(*,\geq 1)}| - 1) \right]^{\frac{1}{\alpha_o}}$$

where the third inequality holds from concavity.

Isolating $\sum_{u \in \mathcal{U}_{b(u)}^{(*,\geq 1)}} d_u w_u$ from the above yields

$$\sum_{u \in \mathcal{U}_{b(u)}^{(*,\geq 1)}} d_u w_u \geq (\eta)^{\overline{\alpha_o}} (|\mathcal{U}_{b(u)}^{(*,\geq 1)}| - 1)^{\frac{\overline{\alpha_o}}{\alpha_o}} \quad (10)$$

Applying the same reasoning to set $\mathcal{U}_{b(u)}^{(*,<1)}$, we obtain

$$\sum_{u \in \mathcal{U}_{b(u)}^{(*,\geq 1)}} d_u w_u \geq (\eta)^{\overline{\alpha'_o}} (|\mathcal{U}_{b(u)}^{(*,\leq 1)}| - 1)^{\frac{\overline{\alpha'_o}}{\alpha_o}} \quad (11)$$

Putting together (10) and (11) yields

$$l_{b(u)} = \sum_{u \in \mathcal{U}_{b(u)}^{(*,\geq 1)}} d_u w_u + \sum_{u \in \mathcal{U}_{b(u)}^{(*,<1)}} d_u w_u$$

$$\geq (\eta)^{\overline{\alpha_o}} (|\mathcal{U}_{b(u)}^{(*,\geq 1)}| - 1)^{\frac{\overline{\alpha_o}}{\alpha_o}} + (\eta)^{\overline{\alpha'_o}} (|\mathcal{U}_{b(u)}^{(*,\leq 1)}| - 1)^{\frac{\overline{\alpha'_o}}{\alpha_o}}$$

The above gives a lower bound for $l_b$ as long there are at least two slices with $\alpha_o \geq 1$ or $\alpha_o < 1$ in base station $b$. We next deal with the case in which there are two slices in base station $b$, $o$ and $o'$, one with $\alpha_o < 1$ and the other with $\alpha_{o'} \geq 1$. It follows from the previous analysis that for this case we have

$$w_u \geq \frac{s_o}{2m_u^o |\mathcal{U}^o|} \frac{(d_v w_v)^{\frac{1}{\alpha_o}}}{(l_b)^{\frac{2}{\alpha_o}-1}} \quad (12)$$

and

$$w_v \geq \frac{s_{o'}}{2m_v^{o'} |\mathcal{U}^{o'}|} \frac{(d_u w_u)^{\frac{1}{\alpha_{o'}}}}{(l_b)^{\frac{2}{\alpha_{o'}}-1}} \quad (13)$$

Combining (12) and (13) we obtain

$$w_u^{\alpha_o} \geq \left( \frac{s_o}{2m_u^o |\mathcal{U}^o|} \right)^{\alpha_o} \frac{d_v}{(l_b)^{2-\alpha_o}} \frac{s_{o'}}{2m_v^{o'} |\mathcal{U}^{o'}|} \frac{(d_u w_u)^{\frac{1}{\alpha_{o'}}}}{(l_b)^{\frac{2}{\alpha_{o'}}-1}}$$

from which

$$l_b^{1-\alpha_o+2/\alpha_{o'}} (w_u)^{\alpha_o - \frac{1}{\alpha_{o'}}} \geq \left( \frac{s_o}{2m_u^o |\mathcal{U}^o|} \right)^{\alpha_o} \frac{d_v (d_u)^{\frac{1}{\alpha_{o'}}} s_{o'}}{2m_v^{o'} |\mathcal{U}^{o'}|}$$

and thus

$$l_b^{1+1/\alpha_{o'}} \left( \frac{w_u}{l_b} \right)^{\alpha_o - 1/\alpha_{o'}} \geq \left( \frac{s_o}{2m_u^o |\mathcal{U}^o|} \right)^{\alpha_o} \frac{d_v (d_u)^{\frac{1}{\alpha_{o'}}} s_{o'}}{2m_v^{o'} |\mathcal{U}^{o'}|}$$

Multiplying each side by $d_u^{\alpha_o - 1/\alpha_{o'}}$ we obtain

$$l_b^{1+1/\alpha_{o'}} \left( \frac{d_u w_u}{l_b} \right)^{\alpha_o - 1/\alpha_{o'}} \geq \left( \frac{s_o}{2m_u^o |\mathcal{U}^o|} \right)^{\alpha_o} \frac{d_v (d_u)^{\alpha_o} s_{o'}}{2m_v^{o'} |\mathcal{U}^{o'}|}$$

from which

$$l_b \left( \frac{d_{b(u)}^o}{l_b} \right)^{\alpha_o - 1/\alpha_{o'}} \geq \left( \frac{s_o}{2m_u^o |\mathcal{U}^o|} \right)^{\alpha_o} \frac{d_v (d_u)^{\alpha_o} s_{o'}}{2m_v^{o'} |\mathcal{U}^{o'}|} \doteq \mu_1 \quad (14)$$

By following the same reasoning as above but isolating $w_v$ instead of $w_u$ from (12) and (13), we obtain

$$l_b \left( \frac{d_{b(v)}^o}{l_b} \right)^{\alpha_{o'} - 1/\alpha_o} \geq \left( \frac{s_{o'}}{2m_v^{o'} |\mathcal{U}^{o'}|} \right)^{\alpha_{o'}} \frac{s_o d_u (d_v)^{\alpha_o}}{2m_u^o |\mathcal{U}^o|} \doteq \mu_2 \quad (15)$$

If $\alpha_o - 1/\alpha_{o'} > 0$, it holds $\left( \frac{d_{b(u)}^o}{l_b} \right)^{\alpha_o - 1/\alpha_{o'}} \leq 1$, and then from (14) we obtain

$$l_b \geq \left( \frac{s_o}{2m_u^o |\mathcal{U}^o|} \right)^{\alpha_o} \frac{s_{o'} d_v (d_u)^{\alpha_o}}{2m_v^{o'} |\mathcal{U}^{o'}|}$$

Otherwise (i.e., in the case $\alpha_o - 1/\alpha_{o'} < 0$), by taking the minimum of both sides of (14) and (15), we obtain the following inequality

$$l_b \min \left( \left( \frac{d_{b(u)}^o}{l_b} \right)^{\alpha_o - \frac{1}{\alpha_{o'}}}, \left( \frac{d_{b(v)}^o}{l_b} \right)^{\alpha_{o'} - \frac{1}{\alpha_o}} \right) \geq \min(\mu_1, \mu_2)$$

From the above,

$$l_b \min \left( \frac{1}{\left( \frac{d_{b(u)}^o}{l_b} \right)^{\hat{\alpha}_o}}, \frac{1}{\left( \frac{d_{b(v)}^o}{l_b} \right)^{\hat{\alpha}_o}} \right) \geq \min(\mu_1, \mu_2)$$

where $\hat{\alpha}_o = \max(-\alpha_o + \frac{1}{\alpha_{o'}}, -\alpha_{o'} + \frac{1}{\alpha_o})$. The above is equivalent to

$$l_b \frac{1}{\left( \max \left( \frac{d_{b(u)}^o}{l_b}, \frac{d_{b(v)}^o}{l_b} \right) \right)^{\hat{\alpha}_o}} \geq \min(\mu_1, \mu_2)$$

Noting that either $\frac{d_{b(u)}^o}{l_b} \geq \frac{1}{2}$ or $\frac{d_{b(v)}^o}{l_b} \geq \frac{1}{2}$,

$$l_b \geq \left( \frac{1}{2} \right)^{\hat{\alpha}_o} \min(\mu_1, \mu_2) \doteq \underline{l_b} > 0$$

The above proves that $l_b$ is bounded in the perturbed game. Hereafter, we denote this bound by $\underline{l_b} > 0$.

*Bound for* $w_u$

Let us start with the case $\alpha_o \geq 1$. Combining (9) with the bound for $l_b$ gives

$$w_u \geq \frac{s_o}{2m_u^o |\mathcal{U}^o| (l_b)^{\frac{2}{\alpha_o}-1}} (\underline{l_b} - d_u w_u)^{\frac{1}{\alpha_o}}$$

$$\geq \frac{s_o}{2m_u^o |\mathcal{U}^o| (l_b)^{\frac{2}{\alpha_o}-1}} \left[ (\underline{l_b})^{\frac{1}{\alpha_o}} - (d_u w_u)^{\frac{1}{\alpha_o}} \right]$$

Since $w_u^{\frac{1}{\alpha_o}} \geq w_u$, it follows that

$$w_u^{\frac{1}{\alpha_o}} + \frac{s_o (d_u w_u)^{\frac{1}{\alpha_o}}}{m_u^o |\mathcal{U}^o| (\underline{l_b})^{\frac{2}{\alpha_o}-1}} \geq \frac{s_o (\underline{l_b})^{\frac{1}{\alpha_o}}}{m_u^o |\mathcal{U}^o| (\underline{l_b})^{\frac{2}{\alpha_o}-1}}$$

from which

$$w_u \geq \left( \frac{s_o (l_{b(u)})^{\frac{1}{\alpha_o}}}{m_u^o |\mathcal{U}^o| (\underline{l_{b(u)}})^{\frac{2}{\alpha_o}-1} + s_o (d_u)^{\frac{1}{\alpha_o}}} \right)^{\alpha_o}$$

which provides a lower bound on $w_u$ for this case. We next look at the case $\alpha_o < 1$. Combining again (9) with the bound for $\underline{l_b}$ gives

$$(w_u)^{\alpha_o} \geq \left(\frac{s_o}{2|\mathcal{U}_o|m_u^o}\right)(\underline{l_b} - d_u w_u)$$

which yields

$$(w_u)^{\alpha_o} + (w_u)^{\alpha_o}\frac{s_o d_u}{2|\mathcal{U}_o|m_u^o} \geq (w_u)^{\alpha_o} + w_u\frac{s_o d_u}{2|\mathcal{U}_o|m_u^o}$$
$$\geq \frac{s_o l_{b(u)}}{|\mathcal{U}_o|m_u^o} \geq \frac{s_o \underline{l_b}}{2|\mathcal{U}_o|m_u^o}$$

Isolating $w_u$ from the above equation, we obtain the following lower bound for this case:

$$w_u \geq \left(\frac{s_o \underline{l_b}}{(2|\mathcal{U}_o|m_u^o + s_o d_u)}\right)^{\frac{1}{\alpha_o}} \doteq \delta$$

With the above, we have obtained a lower bound on $w_u$ for all possible cases. Hereafter we denote this lower bound by $\delta$. □

Building on the result of the above lemma, we proceed as follows. Let us consider a decreasing sequence $\varepsilon^k \to 0$, where $\varepsilon^0 \leq \varepsilon_{max}$, and let $\mathbf{w}^{(\varepsilon^k)}$ denote the Nash Equilibrium of game $G^{(\varepsilon^k)}$. Since the strategy space $\mathcal{S} = \mathcal{S}^1 \times ... \times \mathcal{S}^{|\mathcal{O}|}$ is a compact set, there exists a subsequence $\varepsilon^{k_n}$ such that $\varepsilon^{k_n} \to 0$ and $\mathbf{w}^{(\varepsilon^{k_n})} \to \mathbf{w}^{(0)}$. Note that $w_u^{(\varepsilon^{k_n})} \geq \delta$ $\forall u$ and therefore $w_u^{(0)} \geq \delta$ $\forall u$. Now, let us define function

$$\mathbf{g}(\varepsilon, \mathbf{w}) = \mathbf{R}^{(\varepsilon)}(\mathbf{w}) - \mathbf{w}$$

where $\mathbf{R}^{(\varepsilon)}(\mathbf{w})$ is the best response to $\mathbf{w}$ in the game $G^{(\varepsilon)}$. Note that $\mathbf{g}(\varepsilon, \mathbf{w})$ is equal to zero at the NE of the perturbed game $G^{(\varepsilon^k)}$. Furthermore, from the above lemma we have that at this NE it holds $w_u > 0$ $\forall u$ (even as $\varepsilon \to 0$), and from Lemma 4 we further have that this function is continuous and differentiable for $w_u > 0$ $\forall u$. Thus, $\mathbf{g}(\varepsilon^{k_n}, \mathbf{w}^{(\varepsilon^{k_n})}) = \mathbf{0}$ and

$$\lim_{\varepsilon^{k_n} \to 0} \mathbf{g}(\varepsilon^{k_n}, \mathbf{w}^{(\varepsilon^{k_n})}) = \mathbf{g}(0, \mathbf{w}^{(0)}) = \mathbf{0}$$

implying that indeed $\mathbf{w}^{(0)}$ exists and is a Nash Equilibrium of our (non perturbed) game.

### Proof of Lemma 3

We consider the following two cases: (i) $\mathbf{\Delta\omega}(r+1) = \mathbf{0}$ and (ii) $\mathbf{\Delta\omega}(r+1) \neq \mathbf{0}$.

*Case 1:* $\mathbf{\Delta\omega}(r+1) = \mathbf{0}$.

From $\mathbf{\Delta\omega}(r+1) = \mathbf{0}$, we have that the lhs of (3) is equal to 1. Furthermore, from $\mathbf{\Delta\omega}(r) \neq \mathbf{0}$ it follows that the rhs must be strictly greater than 1. The inequality for this case follows from these two results.

*Case 2:* $\mathbf{\Delta\omega}(r+1) \neq \mathbf{0}$.

Let us define $\mathbf{\Delta a}^o(t) = (\Delta a_b^o(t) : b \in \mathcal{B})$ such that

$$a_b^o(t) = a_b^o(t - |\mathcal{O}| + 1)(1 + \Delta a_b^o(t)).$$

Note that, if operator $o$ updates its best response at time $t$, then $a_b^o(t)$ is the congestion that operator $o$ sees upon making its update, and $\Delta a_b^o(t)$ corresponds to the congestion variation relative to its previous update slot.

In order to prove (3), we will begin by showing that for any slice $o$ updating its weights in round $r+1$, say at time $t+1$, the following holds:

$$\max\left(1 + \overline{\Delta}w^o(t+1), \frac{1}{1 + \underline{\Delta}w^o(t+1)}\right) \tag{16}$$
$$< \max_{t' \in \{t-|\mathcal{O}|+1,...,t\}}\left(1 + \overline{\Delta}w^{o(t')}(t'), \frac{1}{1 + \underline{\Delta}w^{o(t')}(t')}\right)$$

where $o(t')$ denotes the slice updating its weights at time $t'$. The above means that the "relative change" in weights for slice $o$ is strictly smaller than that seen by the other slices in the previous $|\mathcal{O}|$ updates.

To prove (16), we will consider two additional subcases: (i) $\mathbf{\Delta a}^o(t) = \mathbf{0}$, and (i) $\mathbf{\Delta a}^o(t) \neq \mathbf{0}$.

*Subcase 2.1:* $\mathbf{\Delta a}^o(t) = \mathbf{0}$.

To prove (16) for this case, we will show that the lhs of (16) is equal to 1 and the rhs is strictly greater than 1. To show that the lhs is equal to 1, we proceed as follows. Note that, as $\mathbf{\Delta a}^o(t) = \mathbf{0}$, the congestion seen by slice $o$ at time $t+1$ is unchanged with respect to its previous update, and therefore $\mathbf{\Delta w}^o(t+1) = \mathbf{0}$. As a result of this, the lhs of (16) is equal to 1.

The proof that the rhs of (16) is strictly greater than 1 follows by contradiction. Suppose that the rhs is equal to 1. This implies that $\mathbf{\Delta w}^{o(t')}(t') = \mathbf{0}$ for $t' \in \{t-|\mathcal{O}|+1,...,t\}$, where $o(t')$ denotes the slice that changes its weights at time $t'$. This yields $\mathbf{\Delta a}^{o(t+2)}(t+1) = \mathbf{0}$, since there have not been any changes in the weights since its last update, which in turn leads to $\mathbf{\Delta w}^{o(t+2)}(t+2) = \mathbf{0}$. Applying this argument recursively, it can be shown that $\mathbf{\Delta w}^{o(t'')}(t') = 0$ for any $t'' > t+1$. This in turn implies that there is now weight change in round $r+1$, i.e., $\mathbf{\Delta\omega}(r+1) = \mathbf{0}$, which contradicts the fact that we are looking at case $\mathbf{\Delta\omega}(r+1) \neq \mathbf{0}$.

*Subcase 2.2:* $\mathbf{\Delta a}^o(t) \neq \mathbf{0}$.

In order to show (16) for this case, we will start by proving the following intermediate result

$$\max\left(1 + \overline{\Delta}w^o(t+1), \frac{1}{1 + \underline{\Delta}w^o(t+1)}\right) <$$
$$\max\left(1 + \overline{\Delta}a^o(t), \frac{1}{1 + \underline{\Delta}a^o(t)}\right), \tag{17}$$

where $\overline{\Delta}a^o(t) = \max_{b \in \mathcal{B}} \Delta a^o(t)$ and $\underline{\Delta}a^o(t) = \min_{b \in \mathcal{B}} \Delta a^o(t)$.[10]

---

[10]Note that we are defining $\overline{\Delta}a^o(t)$ and $\underline{\Delta}a^o(t)$ across all base stations, and not only those where slice $o$ has users; for the base stations where slice $o$ does not have users, $a^o(t)$ will simply be the load of the base station.

If $\mathbf{\Delta w}^o(t+1) = \mathbf{0}$, (17) follows from the fact that the lhs of (17) is equal to 1 and the rhs is strictly greater than 1 (given that $\mathbf{\Delta a}^o(t) \neq \mathbf{0}$).

We next consider the case where $\mathbf{\Delta w}^o(t+1) \neq \mathbf{0}$. For this case, we prove (17) by showing that the each of two terms of the lhs of (17) is strictly lower than the rhs. First, we show this for the first term, i.e.,

$$\left(1 + \overline{\Delta} w^o(t+1)\right) < \max\left(1 + \overline{\Delta} a^o(t), \frac{1}{1 + \underline{\Delta} a^o(t)}\right). \quad (18)$$

We prove (18) by contradiction: we suppose that

$$1 + \overline{\Delta} w^o(t+1) \geq \max\left(1 + \overline{\Delta} a^o(t), \frac{1}{1 + \underline{\Delta} a^o(t)}\right). \quad (19)$$

and show that this yields a contradiction.

Let $u, v \in \mathcal{U}^o$ be, respectively, the user for which $\Delta w_u(t+1)$ takes the largest value and the one for which it takes the smallest value. Then,

$$\frac{w_u(t+1)}{w_v(t+1)} = \frac{w_u(t)(1 + \overline{\Delta} w^o(t+1))}{w_v(t)(1 + \underline{\Delta} w^o(t+1))} \quad (20)$$

where $\overline{\Delta} w^o(t+1) = \Delta w_u(t+1)$ and $\underline{\Delta} w^o(t+1) = \Delta w_v(t+1)$

For readability purposes, let us denote the base station of user $u$ by $b$ and the base station of user $v$ by $b'$. From (2), we have

$$\frac{w_u(t+1)}{w_v(t+1)} = \frac{\frac{\beta_u \left(a_b^o(t-|\mathcal{O}|+1)(1+\Delta a_b^o(t))\right)^{\frac{1}{\alpha_o}}}{\left(a_b^o(t-|\mathcal{O}|+1)(1+\Delta a_b^o(t))+d_u w_u(t)(1+\Delta w_u(t+1))\right)^{\frac{2}{\alpha_o}-1}}}{\frac{\beta_v \left(a_{b'}^o(t-|\mathcal{O}|+1)(1+\Delta a_{b'}^o(t))\right)^{\frac{1}{\alpha_o}}}{\left(a_{b'}^o(t-|\mathcal{O}|+1)(1+\Delta a_{b'}^o(t))+d_v w_v(t)(1+\Delta w_v(t+1))\right)^{\frac{2}{\alpha_o}-1}}}$$

$$= \frac{\frac{\beta_u \left(a_b^o(t-|\mathcal{O}|+1)(1+\Delta a_b^o(t))\right)^{\frac{1}{\alpha_o}}}{\left(a_b^o(t-|\mathcal{O}|+1)(1+\Delta a_b^o(t))+d_u w_u(t)(1+\overline{\Delta} w^o(t+1))\right)^{\frac{2}{\alpha_o}-1}}}{\frac{\beta_v \left(a_{b'}^o(t-|\mathcal{O}|+1)(1+\Delta a_{b'}^o(t))\right)^{\frac{1}{\alpha_o}}}{\left(a_{b'}^o(t-|\mathcal{O}|+1)(1+\Delta a_{b'}^o(t))+d_v w_v(t)(1+\underline{\Delta} w^o(t+1))\right)^{\frac{2}{\alpha_o}-1}}}$$

By taking the largest and smallest values in $\mathbf{\Delta a}^o(t)$, we obtain the following inequality:

$$\frac{w_u(t+1)}{w_v(t+1)} \leq \frac{\beta_u}{\beta_v} \frac{\frac{\left(a_b^o(t-|\mathcal{O}|+1)(1+\overline{\Delta} a^o(t))\right)^{\frac{1}{\alpha_o}}}{\left(a_b^o(t-|\mathcal{O}|+1)(1+\overline{\Delta} a^o(t))+d_u w_u(t)(1+\overline{\Delta} w^o(t+1))\right)^{\frac{2}{\alpha_o}-1}}}{\frac{\left(a_{b'}^o(t-|\mathcal{O}|+1)(1+\underline{\Delta} a^o(t))\right)^{\frac{1}{\alpha_o}}}{\left(a_{b'}^o(t-|\mathcal{O}|+1)(1+\underline{\Delta} a^o(t))+d_v w_v(t)(1+\underline{\Delta} w^o(t+1))\right)^{\frac{2}{\alpha_o}-1}}}$$

$$= \frac{\beta_u}{\beta_v} \times$$

$$\frac{\frac{\left(a_b^o(t-|\mathcal{O}|+1)\right)^{\frac{1}{\alpha_o}}}{\left(a_b^o(t-|\mathcal{O}|+1)\frac{(1+\overline{\Delta} a^o(t))}{(1+\overline{\Delta} a^o(t))^{\frac{1}{2-\alpha_o}}}+d_u w_u(t)\frac{(1+\overline{\Delta} w^o(t+1))}{(1+\overline{\Delta} a^o(t))^{\frac{1}{2-\alpha_o}}}\right)^{\frac{2}{\alpha_o}-1}}}{\frac{\left(a_{b'}^o(t-|\mathcal{O}|+1)(1+\underline{\Delta} a^o(t))\right)^{\frac{1}{\alpha_o}}}{\left(a_{b'}^o(t-|\mathcal{O}|+1)(1+\underline{\Delta} a^o(t))+d_v w_v(t)(1+\underline{\Delta} w^o(t+1))\right)^{\frac{2}{\alpha_o}-1}}}$$

From our assumption, we have that $1 + \overline{\Delta} w^o(t+1) \geq 1 + \overline{\Delta} a^o(t)$, from which it follows that:

$$\frac{w_u(t+1)}{w_v(t+1)} \leq \frac{\frac{\beta_u \left(a_b^o(t-|\mathcal{O}|+1)\right)^{\frac{1}{\alpha_o}}}{\left(a_b^o(t-|\mathcal{O}|+1)+d_u w_u(t)\right)^{\frac{2}{\alpha_o}-1}} \left(\frac{(1+\overline{\Delta} a^o(t))}{(1+\overline{\Delta} a^o(t))^{\frac{1}{2-\alpha_o}}}\right)^{\frac{2}{\alpha_o}-1}}{\frac{\beta_v \left(a_{b'}^o(t-|\mathcal{O}|+1)(1+\underline{\Delta} a^o(t))\right)^{\frac{1}{\alpha_o}}}{\left(a_{b'}^o(t-|\mathcal{O}|+1)(1+\underline{\Delta} a^o(t))+d_v w_v(t)(1+\underline{\Delta} w^o(t+1))\right)^{\frac{2}{\alpha_o}-1}}}$$

$$\leq \frac{\frac{\beta_u \left(a_b^o(t-|\mathcal{O}|+1)\right)^{\frac{1}{\alpha_o}} (1+\overline{\Delta} a^o(t))^{1-\frac{1}{\alpha_o}}}{\left(a_b^o(t-|\mathcal{O}|+1)+d_u w_u(t)\right)^{\frac{2}{\alpha_o}-1}}}{\frac{\beta_v \left(a_{b'}^o(t-|\mathcal{O}|+1)(1+\underline{\Delta} a^o(t))\right)^{\frac{1}{\alpha_o}}}{\left(a_{b'}^o(t-|\mathcal{O}|+1)(1+\underline{\Delta} a^o(t))+d_v w_v(t)(1+\underline{\Delta} w^o(t+1))\right)^{\frac{2}{\alpha_o}-1}}}$$

$$< \frac{\frac{\beta_u \left(a_b^o(t-|\mathcal{O}|+1)\right)^{\frac{1}{\alpha_o}} (1+\overline{\Delta} a^o(t))^{1-\frac{1}{\alpha_o}}}{\left(a_b^o(t-|\mathcal{O}|+1)+d_u w_u(t)\right)^{\frac{2}{\alpha_o}-1}}}{\frac{\beta_v \left(a_{b'}^o(t-|\mathcal{O}|+1)(1+\underline{\Delta} a^o(t))\right)^{\frac{1}{\alpha_o}}}{\left(a_{b'}^o(t-|\mathcal{O}|+1)+d_v w_v(t)\right)^{\frac{2}{\alpha_o}-1}}}$$

$$= \frac{\frac{\beta_u \left(a_b^o(t-|\mathcal{O}|+1)\right)^{\frac{1}{\alpha_o}}}{\left(a_b^o(t-|\mathcal{O}|+1)+d_u w_u(t)\right)^{\frac{2}{\alpha_o}-1}}}{\frac{\beta_v \left(a_{b'}^o(t-|\mathcal{O}|+1)\right)^{\frac{1}{\alpha_o}}}{\left(a_{b'}^o(t-|\mathcal{O}|+1)+d_v w_v(t)\right)^{\frac{2}{\alpha_o}-1}}} \frac{(1+\overline{\Delta} a^o(t))^{1-\frac{1}{\alpha_o}}}{(1+\underline{\Delta} a^o(t))^{\frac{1}{\alpha_o}}}$$

$$= \frac{w_u(t)}{w_v(t)} \frac{(1+\overline{\Delta} a^o(t))^{1-\frac{1}{\alpha_o}}}{(1+\underline{\Delta} a^o(t))^{\frac{1}{\alpha_o}}}.$$

where the third step holds since $\underline{\Delta} a^o(t) < 0$ and $\underline{\Delta} w^o(t+1) < 0$, and the last one follows from the fact that $w_u(t) = w_u(t - |\mathcal{O}| + 1)$ (as slice $o$ does not updated its weights in the time interval $\{t - |\mathcal{O}| + 2, ..., t\}$). Combining (20) with the above yields

$$\frac{w_u(t)(1+\overline{\Delta} w^o(t+1))}{w_v(t)(1+\underline{\Delta} w^o(t+1))} < \frac{w_u(t)}{w_v(t)} \frac{(1+\overline{\Delta} a^o(t))^{1-\frac{1}{\alpha_o}}}{(1+\underline{\Delta} a^o(t))^{\frac{1}{\alpha_o}}}.$$

From the fact that $x^a y^b \leq \max(x, y)$ for $x, y \geq 1$ and $a + b = 1$ we have

$$\frac{(1+\overline{\Delta} a^o(t+1))^{1-\frac{1}{\alpha_o}}}{(1+\underline{\Delta} a^o(t))^{\frac{1}{\alpha_o}}} \leq \max\left(1 + \overline{\Delta} a^o(t), \frac{1}{1 + \underline{\Delta} a^o(t)}\right).$$

From the above two equations we have

$$\frac{1+\overline{\Delta} w^o(t+1)}{1+\underline{\Delta} w^o(t+1)} < \max\left(1 + \overline{\Delta} a^o(t), \frac{1}{1 + \underline{\Delta} a^o(t)}\right).$$

and combining this with (19) yields

$$\frac{1}{1 + \underline{\Delta} w^o(t+1)} < 1$$

which contradicts $\frac{1}{1 + \underline{\Delta} w^o(t+1)} > 1$, thus proving (18).

The next step to prove (17) is to show that the second term of the lhs of (17) is strictly lower than the rhs, i.e.,

$$\frac{1}{1 + \underline{\Delta} w^o(t+1)} < \max\left(1 + \overline{\Delta} a^o(t), \frac{1}{1 + \underline{\Delta} a^o(t)}\right). \quad (21)$$

The proof is analogous to the one for (18) and proceeds again by contradiction: we suppose that

$$\frac{1}{1 + \underline{\Delta} w^o(t+1)} \geq \max\left(1 + \overline{\Delta} a^o(t), \frac{1}{1 + \underline{\Delta} a^o(t)}\right), \quad (22)$$

and show that this yields a contradiction.

Employing a similar argument to the one above, we have:

$$\frac{w_u(t+1)}{w_v(t+1)} < \frac{\frac{\beta_u\left(a_b^o(t-|\mathcal{O}|+1)\right)^{\frac{1}{\alpha_o}}}{\left(a_b^o(t-|\mathcal{O}|+1)+d_u w_u(t)\right)^{\frac{2}{\alpha_o}-1}}}{\frac{\beta_v\left(a_{b'}^o(t-|\mathcal{O}|+1)\right)^{\frac{1}{\alpha_o}}}{\left(a_{b'}^o(t-|\mathcal{O}|+1)+d_v w_v(t)\right)^{\frac{2}{\alpha_o}-1}}} \frac{(1+\overline{\Delta}a^o(t+1))^{\frac{1}{\alpha_o}}}{(1+\underline{\Delta}a^o(t))^{1-\frac{1}{\alpha_o}}}$$

$$= \frac{w_u(t)}{w_v(t)} \frac{(1+\overline{\Delta}a^o(t))^{\frac{1}{\alpha_o}}}{(1+\underline{\Delta}a^o(t))^{1-\frac{1}{\alpha_o}}}.$$

Combining (20) with the above yields

$$\frac{w_u(t)(1+\overline{\Delta}w^o(t+1))}{w_v(t)(1+\underline{\Delta}w^o(t+1))} < \frac{w_u(t)}{w_v(t)} \frac{(1+\overline{\Delta}a^o(t))^{\frac{1}{\alpha_o}}}{(1+\underline{\Delta}a^o(t))^{1-\frac{1}{\alpha_o}}}.$$

Again from $x^a y^b \leq \max(x,y)$ for $x, y \geq 1$ and $a+b=1$ we have

$$\frac{(1+\overline{\Delta}a^o(t))^{\frac{1}{\alpha_o}}}{(1+\underline{\Delta}a^o(t))^{1-\frac{1}{\alpha_o}}} \leq \max\left(1+\overline{\Delta}a^o(t), \frac{1}{1+\underline{\Delta}a^o(t)}\right).$$

From the above two equations

$$\frac{1+\overline{\Delta}w^o(t+1)}{1+\underline{\Delta}w^o(t+1)} < \max\left(1+\overline{\Delta}a^o(t), \frac{1}{1+\underline{\Delta}a^o(t)}\right).$$

Combining the above with (22) leads to $(1+\overline{\Delta}w^o(t+1)) > 1$, which contradicts the fact that $\overline{\Delta}w^o(t+1)$ is necessarily strictly greater than 1.

The above proves our intermediate step (17). Next, building on this intermediate result, we shall show that (16) holds.

Recall that $o(t')$ denotes the slice that updates its weights at time slot $t'$. We have that

$$a_b^o(t) = \sum_{u \in \mathcal{U}_b \setminus \mathcal{U}_b^o} w_u(t) = \sum_{u \in \mathcal{U}_b \setminus \mathcal{U}_b^o} w_u(t(u))$$

$$= \sum_{u \in \mathcal{U}_b \setminus \mathcal{U}_b^o} w_u(t(u)-1)(1+\Delta w_u(t(u)))$$

where $t(u)$ represents the time when user $u$ updates its weights within the interval $\{t-|\mathcal{O}|+2, ..., t\}$.

If we take the maximum $\Delta w_u(t(u))$ over all users, we obtain the following inequality

$$a_b^o(t) \leq \max_{u \in \mathcal{U}_b \setminus \mathcal{U}_b^o} (1+\Delta w_u(t(u))) \sum_{u \in \mathcal{U}_b \setminus \mathcal{U}_b^o} w_u(t(u)-1)$$

$$= \max_{u \in \mathcal{U}_b \setminus \mathcal{U}_b^o} (1+\Delta w_u(t(u))) a_b^o(t-|\mathcal{O}|+1).$$

where the second inequality follows from the fact that we are taking the weight values prior to their update.

Given that $a_b^o(t) = a_b^o(t-|\mathcal{O}|+1)(1+\Delta a_b^o(t))$, the above leads to

$$1+\Delta a_b^o(t) \leq \max_{u \in \mathcal{U}_b \setminus \mathcal{U}_b^o} (1+\Delta w_u(t(u)))$$

From the definition of $\overline{\Delta}w^o(t)$ we have that $\overline{\Delta}w^o(t) \geq \Delta w_u(t)$ for $u \in \mathcal{U}^o$, from which

$$1+\Delta a_b^o(t) \leq \max_{t' \in \{t-|\mathcal{O}|+2,...,t\}} 1+\overline{\Delta}w^{o(t')}(t')$$

Since the above inequality holds for all $b \in \mathcal{B}$, it also holds if we take the maximum value over all $b$ in the rhs, which leads to

$$1+\overline{\Delta}a^o(t) \leq \max_{t' \in \{t-|\mathcal{O}|+2,...,t\}} 1+\overline{\Delta}w^{o(t')}(t').$$

Similarly, it can be shown that

$$1+\underline{\Delta}a^o(t) \geq \min_{t' \in \{t-|\mathcal{O}|+2,...,t\}} 1+\underline{\Delta}w^{o(t')}(t').$$

Combining the above with (17) yields

$$\max\left(1+\overline{\Delta}w^o(t+1), \frac{1}{1+\underline{\Delta}w^o(t+1)}\right)$$

$$< \max_{t' \in \{t-|\mathcal{O}|+2,...,t\}} \left(1+\overline{\Delta}w^{o(t')}(t'), \frac{1}{1+\underline{\Delta}w^{o(t')}(t')}\right)$$

If we add the term $t-|\mathcal{O}|+1$ in the max operation of the rhs, the inequality still holds, thus

$$\max\left(1+\overline{\Delta}w^o(t+1), \frac{1}{1+\underline{\Delta}w^o(t+1)}\right)$$

$$< \max_{t' \in \{t-|\mathcal{O}|+1,...,t\}} \left(1+\overline{\Delta}w^{o(t')}(t'), \frac{1}{1+\underline{\Delta}w^{o(t')}(t')}\right)$$

With the above we have proven (16). In the next (and last) step of the proof, we now show (3) building on this result. For conciseness, we define:

$$f(t) = \max\left(1+\overline{\Delta}w^{o(t)}(t), \frac{1}{1+\underline{\Delta}w^{o(t)}(t)}\right). \quad (23)$$

With the above definition, (16) can be rewritten as

$$f(t+1) < \max_{t' \in \{t-|\mathcal{O}|+1,...,t\}} f(t'). \quad (24)$$

Applying the same argument to the update at time slot $t+2$ yields

$$f(t+2) < \max_{t' \in \{t-|\mathcal{O}|+2,...,t+1\}} f(t'),$$

$$= \max\left[\max_{t' \in \{t-|\mathcal{O}|+2,...,t\}} f(t'), f(t+1)\right]$$

and combining the above with (24) we obtain

$$f(t+2) < \max_{t' \in \{t-|\mathcal{O}|+1,...,t\}} f(t').$$

If we now apply the above argument recursively for $t+3, t+4, t+5, \ldots$, we obtain

$$f(t+i) \leq \max_{t' \in \{t-|\mathcal{O}|+1,...,t\}} f(t'), \ i < 1. \quad (25)$$

Without loss of generality, let assume that $t+1$ coincides with the start of a round; then from the above it follows that

$$\max_{i \in \{1,...,|\mathcal{O}|\}} f(t+i) < \max_{j \in \{1,...,|\mathcal{O}|\}} f(t-|\mathcal{O}|+j) \quad (26)$$

where the max on the lhs includes the updates corresponding to round $r+1$, while those on the rhs correspond to round $r$.

Finally, expressing the above equation in terms of $\Delta\omega_u^o(r)$ leads to (3). □

*Proof of Theorem 3*

The slices responses at the end of each round are captured by the sequence $\boldsymbol{\omega}(r)$ for $r = 1, 2, ...$ and the dynamics are given by $\boldsymbol{\omega}(r+1) = \mathbf{R}(\boldsymbol{\omega}(r))$, where $\mathbf{R}(\boldsymbol{\omega}(r))$ is the result of applying sequentially the best response for each of the slices. Note that $\boldsymbol{\omega}(r)$ is in the set $\mathcal{S} = \mathcal{S}^1 \times \mathcal{S}^2 \times ... \times \mathcal{S}^{|\mathcal{O}|}$ where:

$$\mathcal{S}^o = \{\boldsymbol{\omega}^o \in \mathbf{R}^o \mid \boldsymbol{\omega}^o \geq \mathbf{0} \text{ and } \sum_{b \in \mathcal{B}} n_b^o \omega_b^o = s^o\}.$$

Let us define a function:

$$V(\boldsymbol{\omega}(r)) := \max_{o \in \mathcal{O}} \left(1 + \overline{\Delta}\omega^o(r+1), \frac{1}{1 + \underline{\Delta}\omega^o(r+1)}\right) - 1$$

Notice that $V(\boldsymbol{\omega}(r))$ is a function of the relative weight changes at round $r+1$ given the weights at round $r$, $\boldsymbol{\omega}(r)$.

Note that, unless algorithm converged (i.e., $\boldsymbol{\omega}(r+1) = \boldsymbol{\omega}(r)$ and $V(\boldsymbol{\omega}(r)) = 0$), there should at least one slice with a user $u$ for which $\Delta\omega_u^o(r+1) > 0$ and a user $v$ for which $\Delta\omega_v^o(r+1) < 0$. Thus, in this case it must be that $V(\boldsymbol{\omega}(r)) > 0$ and

$$\max_{o \in \mathcal{O}} \left(1 + \overline{\Delta}\omega^o(r+1), \frac{1}{1 + \underline{\Delta}\omega^o(r+1)}\right) > 1$$

Note also that by Lemma 3 it follows that if $\boldsymbol{\omega}(r+1) \neq \boldsymbol{\omega}(r)$, then

$$\max_{o \in \mathcal{O}} \left(1 + \overline{\Delta}\omega^o(r+1), \frac{1}{1 + \underline{\Delta}\omega^o(r+1)}\right) < \max_{o \in \mathcal{O}} \left(1 + \overline{\Delta}\omega^o(r), \frac{1}{1 + \underline{\Delta}\omega^o(r)}\right), \quad (27)$$

which is equivalent to $V(\boldsymbol{\omega}(r+1)))) < V(\boldsymbol{\omega}(r))$.

Note that, since $\underline{\Delta}\omega^o > -1$, $V(\boldsymbol{\omega}(r))$ is continuous. Furthermore, the best response functions governing $\Delta$ are also continuous, as shown in Theorem 2.

In summary, the above results show that $V(\boldsymbol{\omega}(r))$ is continuous, non-negative and decreasing each round. So, we might expect it to converge to $0$ in which case the slices weights must have converged. We show this by contradiction. Suppose $V(\boldsymbol{\omega}(r))$ converges instead to $\epsilon > 0$, so $V(\boldsymbol{\omega}(0)) \geq V(\boldsymbol{\omega}(r)) \geq \epsilon$ for all $r$. Let us define $\mathcal{V} = \{\boldsymbol{\omega} | V(\boldsymbol{\omega}(0)) \geq V(\boldsymbol{\omega}) \geq \epsilon\}$ and let $\mathcal{C} = \{\boldsymbol{\omega} | \boldsymbol{\omega} \in \mathcal{S} \cap \mathcal{V}\}$. From this, $\mathcal{V}$ is closed, and since $\mathcal{S}$ is compact, so is $\mathcal{C}$. From the continuity of $V$ we have that:

$$\delta = \max_{\boldsymbol{\omega} \in C} V(\mathbf{R}(\boldsymbol{\omega}))) - V(\boldsymbol{\omega}) < 0$$

Clearly it is not possible $V(\boldsymbol{\omega}(r)) \geq \epsilon$ since each round it decreases by at least $\delta$. It follows that the weights must converge when $V(\boldsymbol{\omega}(r)) = 0$. □

*Proof of Theorem 4*

The proof follows from the following three lemmas.

**Lemma 6.** *An optimal (not necessarily unique) solution to the centralized problem is given by $\mathbf{w}^*$ which assigns weights to all users of a given slice proportionally to their priorities, i.e., $w_u^* = \phi_u s_o, \ \forall u \in \mathcal{U}_o$.*

*Proof.* We need only show that $U(\mathbf{w}^*) \geq U(\mathbf{w})$ for any other feasible weight vector $\mathbf{w}$. To that end, consider

$$U(\mathbf{w}^*) - U(\mathbf{w}) = \sum_{o \in \mathcal{O}} \sum_{u \in \mathcal{U}_o} \phi_u \left(\log\left(\frac{w_u^* c_u}{l_b(\mathbf{w}^*)}\right) - \log\left(\frac{w_u c_u}{l_b(\mathbf{w})}\right)\right)$$

Let us denote the distributions induced by $\mathbf{w}^*$ and $\mathbf{w}$ respectively as: $\mathbf{p}^b(\mathbf{w}) = (p_u^b(\mathbf{w}) = \frac{w_u}{l_b(\mathbf{w})} : u \in \mathcal{U}_b)$ and $\mathbf{p}^b(\mathbf{w}^*) = (p_u^b(\mathbf{w}^*) = \frac{w_u}{l_b(\mathbf{w}^*)} : u \in \mathcal{U}_b)$. Since $\boldsymbol{\phi} = \mathbf{w}^*$, we have

$$U(\mathbf{w}^*) - U(\mathbf{w}) = \sum_{b \in \mathcal{B}} l_b(\mathbf{w}^*) \sum_{o \in \mathcal{O}} \sum_{u \in \mathcal{U}_o^b} p_u^b(w^*) \log\left(\frac{p_u^b(w^*)}{p_u^b(w)}\right)$$

$$= \sum_{b \in \mathcal{B}} l_b(\mathbf{w}^*) D(\mathbf{p}^b(\mathbf{w}^*) || \mathbf{p}^b(\mathbf{w}))$$

where $D(\mathbf{p}^b(\mathbf{w}^*) || \mathbf{p}^b(\mathbf{w}))$ is the Kullback-Leibler divergence, between the distributions induced by $\mathbf{w}^*$ and $\mathbf{w}$ respectively, i.e., $\mathbf{p}^b(\mathbf{w}^*)$ and $\mathbf{p}^b(\mathbf{w})$. It is known [25] that $D(\mathbf{p}^b(\mathbf{w}^*) || \mathbf{p}^b(\mathbf{w})) \geq 0$ and $0$ only when $\mathbf{p}^b(\mathbf{w}) = \mathbf{p}^b(\mathbf{w}^*)$ Hence it follows that $\mathbf{w}^*$ is optimal. □

**Lemma 7.** *Let $\mathbf{w}$ be a Nash Equilibrium of the distributed resource allocation game, and $\mathbf{w}^*$ an optimal solution. Then, $U(\mathbf{w}^*) - U(\mathbf{w}) \leq \log(e)$.*

*Proof.* Since in the Nash Equilibrium each slice maximizes its utility given the allocation of the other slices,

$$\sum_{u \in \mathcal{U}^o} \phi_u \log\left(\frac{w_u}{l_{b(u)}(\mathbf{w})}\right) \geq \sum_{u \in \mathcal{U}^o} \phi_u \log\left(\frac{w_u^*}{d_{b(u)}^o(\mathbf{w}^*) + a_{b(u)}^o(\mathbf{w})}\right)$$

Given that $d_{b(u)}^o(\mathbf{w}^*) + a_{b(u)}^o(\mathbf{w}) \leq l_{b(u)}(\mathbf{w}) + l_{b(u)}(\mathbf{w}^*)$,

$$\sum_{u \in \mathcal{U}^o} \phi_u \log\left(\frac{w_u}{l_{b(u)}(\mathbf{w})}\right) \geq \sum_{u \in \mathcal{U}^o} \phi_u \log\left(\frac{w_u^*}{l_{b(u)}(\mathbf{w}) + l_{b(u)}(\mathbf{w}^*)}\right)$$

From the above it follows that

$$\sum_{u \in \mathcal{U}^o} \phi_u \log(r_u(\mathbf{w}^*)) - \sum_{u \in \mathcal{U}^o} \phi_u \log(r_u(\mathbf{w}))$$

$$\leq \sum_{u \in \mathcal{U}^o} \phi_u \log\left(\frac{w_u^* c_u}{l_{b(u)}(\mathbf{w}^*)}\right) - \sum_{u \in \mathcal{U}^o} \phi_u \log\left(\frac{w_u^* c_u}{l_{b(u)}(\mathbf{w}) + l_{b(u)}(\mathbf{w}^*)}\right)$$

$$= -\sum_{u \in \mathcal{U}^o} \phi_u \log\left(\frac{l_{b(u)}(\mathbf{w}^*)}{l_{b(u)}(\mathbf{w}) + l_{b(u)}(\mathbf{w}^*)}\right)$$

Summing the above over all slices weighted by the corresponding shares yields

$$U(\mathbf{w}^*) - U(\mathbf{w}) \leq -\sum_{u \in \mathcal{U}} \phi_u s_o \log\left(\frac{l_{b(u)}(\mathbf{w}^*)}{l_{b(u)}(\mathbf{w}) + l_{b(u)}(\mathbf{w}^*)}\right)$$

Given $w_u^* = \phi_u s_o$, we have

$$U(\mathbf{w}^*) - U(\mathbf{w}) \leq -\sum_{b \in \mathcal{B}} \log\left(\frac{l_b(\mathbf{w}^*)}{l_b(\mathbf{w}) + l_b(\mathbf{w}^*)}\right)^{\sum_{u \in \mathcal{U}_b} w_u^*}$$

$$= -\sum_{b \in \mathcal{B}} \sum_{u \in \mathcal{U}_b} w_u \log \left( \frac{l_b(\mathbf{w}^*)}{l_b(\mathbf{w}) + l_b(\mathbf{w}^*)} \right)^{\frac{\sum_{v \in \mathcal{U}_b} w_v^*}{\sum_{v \in \mathcal{U}_b} w_v}}$$

$$= -\sum_{b \in \mathcal{B}} \sum_{u \in \mathcal{U}_b} w_u \log \left( \frac{l_b(\mathbf{w}^*)/l_b(\mathbf{w})}{1 + l_b(\mathbf{w}^*)/l_b(\mathbf{w})} \right)^{\frac{l_b(\mathbf{w}^*)}{l_b(\mathbf{w})}}$$

and, given that $(x/(1+x))^x > 1/e$ for $x \geq 0$, this yields

$$U(\mathbf{w}^*) - U(\mathbf{w}) \leq \sum_{b \in \mathcal{B}} \sum_{u \in \mathcal{U}_b} w_u \log(e) = \log(e). \qquad \square$$

**Lemma 8.** *There exists some scenario for which $U(\mathbf{w}^*) - U(\mathbf{w}) = \log(e)$.*

*Proof.* Let us consider a scenario with two slices with shares $s_1$ and $s_2$, respectively. There are two base stations. Slice 1 has $m+1$ users, $m$ associated to base station 1 and one associated to base station 2. Slice 2 has one user associated to base station 2. All users have $c_{ub} = 1$. Under the optimal allocation:

$$U(\mathbf{w}^*) = \frac{s_1}{m+1} m \log\left(\frac{1}{m}\right) + \frac{s_1}{m+1} \log\left(\frac{\frac{s_1}{m+1}}{\frac{s_1}{(m+1)} + s_2}\right)$$

$$+ s_2 \log\left(\frac{s_2}{\frac{s_1}{(m+1)} + s_2}\right),$$

and under the Nash equilibrium

$$U(\mathbf{w}) = \frac{s_1}{m+1} m \log\left(\frac{1}{m}\right) + \frac{s_1}{m+1} \log\left(\frac{s_1}{s_1 + s_2}\right)$$

$$+ s_2 \log\left(\frac{s_2}{s_1 + s_2}\right).$$

For $m \to \infty$ this yields $U(\mathbf{w}^*) = s_1 \log\left(\frac{1}{m}\right) + s_2 \log(1)$ and $U(\mathbf{w}) = s_1 \log\left(\frac{1}{m}\right) + s_2 \log\left(\frac{s_2}{s_1 + s_2}\right)$. From this,

$$U(\mathbf{w}) - U(\mathbf{w}^*) = s_2 \log\left(\frac{s_2}{s_1 + s_2}\right)$$

which tends to $-\log(e)$ when $s_1 \to 1$ and $s_2 \to 0$. $\qquad \square$

*Proof of Theorem 5*

**Lemma 9.** *Let us consider two slices, $o$ and $o'$, that have the same share $s_o$. Let the utility function of slice $o$ be $U^o = \sum_{u \in \mathcal{U}^o} \phi_u \log(r_u)$. Then it holds*

$$U^o(\tilde{\mathbf{w}}_\mathbf{o}) - U^o(\mathbf{w}_\mathbf{o}) \leq 0.060$$

*Proof.* In order to bound the envy $U^o(\tilde{\mathbf{w}}) - U^o(\mathbf{w})$ at the NE, we will construct a weight allocation $\mathbf{m}$ that satisfies $U^o(\mathbf{m}) \leq U^o(\mathbf{w})$ and $U^o(\tilde{\mathbf{m}}) \geq U^o(\tilde{\mathbf{w}})$ – where $\tilde{\mathbf{w}}$ and $\tilde{\mathbf{m}}$ are the allocations resulting from exchanging the resources of slices $o$ and $o'$ in $\mathbf{w}$ and $\mathbf{m}$, respectively. It then follows that $U^o(\tilde{\mathbf{m}}) - U^o(\mathbf{m})$ is an upper bound on the envy.

Specifically, the weight allocation $\mathbf{m}$ will be chosen such that: (*i*) for all slices different from $o$, the weights remain the same as in the NE, i.e, $\mathbf{m}^{-o} = \mathbf{w}^{-o}$; and (*ii*) the weights of slice $o$ are chosen so as to maximize $U^o(\mathbf{m})$ subject to $d_b^o(\mathbf{m}^o) = \sum_{u \in \mathcal{U}_b^o} m_u \leq a_b^o(\mathbf{m}^{-o}) \; \forall b \in \mathcal{B}$ and slice o's share constraint. Note that with this weight allocation we have $a_{b(u)}^o(\mathbf{m}^{-o}) = a_{b(u)}^o(\mathbf{w}^{-o})$ – for readability purposes, we will use just $a_{b(u)}^o$. Note also that the weights that slice $o$ would have with the resources of $o'$ remain the same, i.e. $\tilde{\mathbf{m}}^o = \tilde{\mathbf{w}}^o$.

By following a similar argument to that of Lemma 2, it can be seen that the above leads to the weights $m_u$ for $u \in \mathcal{U}^o$ solving the set of equations below, which have a feasible solution as long as $s_o < \sum_{u \in \mathcal{U}^o} a_{b(u)}^o(\mathbf{m}^{-o})$ (we deal with the case $\sum_{b \in \mathcal{B}_o} a_b^o < s_o$ later).

$$m_u = \begin{cases} a_{b(u)}^o \dfrac{\phi_u}{\sum_{v \in \mathcal{U}_{b(u)}^o} \phi_v}, & a_{b(u)}^o = d_{b(u)}^o(\mathbf{m}^o) \\[1em] \dfrac{\phi_u \dfrac{a_{b(u)}^o}{a_{b(u)}^o + d_{b(u)}^o(\mathbf{m}^o)}}{\sum_{v \in \hat{\mathcal{U}}^o} \phi_v \dfrac{a_{b(v)}^o}{a_{b(v)}^o + d_{b(v)}^o(\mathbf{m}^o)}} s'_o, & a_{b(u)}^o > d_{b(u)}^o(\mathbf{m}^o) \end{cases}$$

where $\hat{\mathcal{U}}^o$ is the set of users of slice $o$ for which $a_{b(u)}^o > d_{b(u)}^o(\mathbf{m}^o)$ and $s'_o = s_o - \sum_{u \in \mathcal{U}^o \setminus \hat{\mathcal{U}}^o} m_u$.

It is clear that with this weight allocation we have $U^o(\mathbf{m}) \leq U^o(\mathbf{w})$. Indeed, only the weights of slice $o$ have changed and (as mentioned before) $\mathbf{w}^o$ is the best response of the slice $o$, hence any other weight setting for this slice will provide a lower utility.

To show $U^o(\tilde{\mathbf{m}}) \geq U^o(\tilde{\mathbf{w}})$ we proceed as follows. The base stations that initially had a load for operator $o$ larger than $a_b^o$ ($d_{b(u)}^o(\mathbf{m}^o) > a_b^o$) decrease their load with the new allocation, while the others increase it. Let us denote the first set of base stations as $\mathcal{B}_1$ and the other set as $\mathcal{B}_2$. Since the base stations of set $\mathcal{B}_1$ decrease their load in the new allocation and the base stations of set $\mathcal{B}_2$ increase it, we can move from the initial allocation to the new one by iteratively selecting one base station of set $\mathcal{B}_1$ and one of set $\mathcal{B}_2$ and moving load from the first one to the second until one of them reaches its target load. When decreasing the load of base station $b$ and increasing that of base station $b'$ by $\delta$ we have

$$\frac{dU^o(\tilde{\mathbf{w}})}{d\delta} = -\frac{\sum_{u \in \mathcal{U}_{b'}^o} \phi_u}{l_{b'}(\tilde{\mathbf{w}})} + \frac{\sum_{u \in \mathcal{U}_b^o} \phi_u}{l_b(\tilde{\mathbf{w}})}$$

If we can show at the beginning (before increasing/decreasing the load of any base station), for any $b \in \mathcal{B}_1$ and $b' \in \mathcal{B}_2$ it holds

$$\frac{\sum_{u \in \mathcal{U}_b^o} \phi_u}{l_b(\tilde{\mathbf{w}})} \geq \frac{\sum_{u \in \mathcal{U}_{b'}^o} \phi_u}{l_{b'}(\tilde{\mathbf{w}})} \qquad (28)$$

we will have the value of $\sum_{u \in \mathcal{U}_b^o} \frac{\phi_u}{l_b}$ for any base station of set $\mathcal{B}_1$ will always be larger than for any base station of set $\mathcal{B}_2$, since it are larger at the beginning and it increases in the intermediate steps, while it decreases for a base station of $\mathcal{B}_2$. With this, $dU^o(\tilde{\mathbf{m}})/d\delta$ is positive at the beginning and will continue to be positive in the intermediate steps, yielding to an increase in $dU^o(\tilde{\mathbf{m}})$.

To show (28), we proceed as follows. It holds that

$$\frac{d_b^o(\mathbf{m}^o)}{d_{b'}^o} = \frac{\sum_{u \in \mathcal{U}_b^o} \phi_u \frac{1}{1 + d_b^o(\mathbf{m}^o)/a_b^o}}{\sum_{u' \in \mathcal{U}_{b'}^o} \phi_{u'} \frac{1}{1 + d_{b'}^o(\mathbf{m}^o)/a_{b'}^o}}$$

$$= \frac{\sum_{u\in\mathcal{U}_b^o}\phi_u}{\sum_{u'\in\mathcal{U}_{b'}^o}\phi_{u'}}\left(\frac{1+\frac{d_{b'}^o(\mathbf{m}^o)}{a_{b'}^o}}{1+\frac{d_b^o(\mathbf{m}^o)}{a_b^o}}\right)$$

For $b\in\mathcal{B}_1$ and $b'\in\mathcal{B}_2$ (since $a_b^o < d_b^o(\mathbf{m}^o)$ and $a_{b'}^o > d_{b'}^o(\mathbf{m}^o)$)

$$\frac{d_b^o(\mathbf{m}^o)}{d_{b'}^o(\mathbf{m}^o)} < \frac{\sum_{u\in\mathcal{U}_b^o}\phi_u}{\sum_{u'\in\mathcal{U}_{b'}^o}\phi_{u'}}$$

and thus

$$\frac{l_b}{\sum_{u\in\mathcal{U}_b^o}\phi_u} = \frac{a_b^o+d_b^o(\mathbf{m}^o)}{\sum_{u\in\mathcal{U}_b^o}\phi_u} < \frac{2d_b^o(\mathbf{m}^o)}{\sum_{u\in\mathcal{U}_b^o}\phi_u} < \frac{2d_{b'}^o(\mathbf{m}^o)}{\sum_{u\in\mathcal{U}_{b'}^o}\phi_u}$$
$$\leq \frac{a_{b'}^o+d_{b'}^o(\mathbf{m}^o)}{\sum_{u\in\mathcal{U}_{b'}^o}\phi_u} = \frac{l_{b'}}{\sum_{u\in\mathcal{U}_{b'}^o}\phi_u}$$

which proves (28), and thus $U^o(\tilde{\mathbf{m}}) \geq U^o(\tilde{\mathbf{w}})$.

We now go back to the case $\sum_{b\in\mathcal{B}_o} a_b^o < s_o$. Following the above procedure, in this case we can find an allocation $\mathbf{m}^o$ that satisfies: (i) $U^o(\mathbf{m}) \leq U^o(\mathbf{w})$, (ii) $U^o(\tilde{\mathbf{m}}) \geq U^o(\tilde{\mathbf{w}})$ and (iii) $d_b^o(\mathbf{m}^o) \geq a_b^o$ $\forall b$. In this case we then have $U^o(\tilde{\mathbf{w}}) - U^o(\mathbf{w}) \leq U^o(\tilde{\mathbf{m}}) - U^o(\mathbf{m}) \leq 0$.

To find an upper bound on $U^o(\tilde{\mathbf{m}}) - U^o(\mathbf{m})$, recall that

$$U^o(\tilde{\mathbf{m}}) = \sum_{u\in\mathcal{U}^o}\phi_u\log\left(\frac{\tilde{m}_u c_u}{l_b(\tilde{\mathbf{m}})}\right),$$

and

$$U^o(\mathbf{m}) = \sum_{u\in\mathcal{U}^o}\phi_u\log\left(\frac{m_u c_u}{l_b(\mathbf{m})}\right).$$

Given that $l_b(\tilde{\mathbf{m}}) = l_b(\mathbf{m})$ and $\tilde{m}_u = m_u$ for $u \notin \hat{\mathcal{U}}^o$, this yields

$$U^o(\tilde{\mathbf{m}}) - U^o(\mathbf{m}) = \sum_{u\in\hat{\mathcal{U}}^o}\phi_u\log(\tilde{m}_u) - \sum_{u\in\hat{\mathcal{U}}^o}\phi_u\log(m_u).$$

Since $\sum_{u\in\hat{\mathcal{U}}^o}\log(\tilde{m}_u)$ subject to $\sum_{u\in\hat{\mathcal{U}}^o}\tilde{m}_u = s'_o$ takes a maximum at $\tilde{m}_u = \hat{\phi}_u s'_o$ (where $\hat{\phi}_u = \phi_u/\sum_{v\in\hat{\mathcal{U}}^o}\phi_v$),

$$U^o(\tilde{\mathbf{m}}) - U^o(\mathbf{m}) \leq \sum_{u\in\hat{\mathcal{U}}^o}\phi_u\log(\hat{\phi}_u s'_o) - \sum_{u\in\hat{\mathcal{U}}^o}\phi_u\log(m_u)$$
$$\leq \sum_{u\in\hat{\mathcal{U}}^o}\hat{\phi}_u\log(\hat{\phi}_u s'_o) - \sum_{u\in\hat{\mathcal{U}}^o}\hat{\phi}_u\log(m_u) \quad (29)$$

In order to bound the term $\sum_{u\in\hat{\mathcal{U}}^o}\hat{\phi}_u\log(m_u)$ above, we look for a bound on $\frac{m_u}{m_v}$. Given that $a_b^o \geq d_b^o(\mathbf{m}^o)$ holds for all $b$, we have for $u,v\in\hat{\mathcal{U}}^o$:

$$\frac{m_u}{m_v} = \frac{\phi_u}{\phi_v}\frac{\frac{a_{b(u)}^o}{a_{b(u)}^o+d_{b(u)}^o(\mathbf{m}^o)}}{\frac{a_{b(v)}^o}{a_{b(v)}^o+d_{b(v)}^o(\mathbf{m}^o)}} > \frac{\phi_u}{\phi_v}\frac{\frac{a_{b(u)}^o}{a_{b(u)}^o+a_{b(u)}^o}}{\frac{a_{b(v)}^o}{a_{b(v)}^o}} = \frac{1}{2}\frac{\hat{\phi}_u}{\hat{\phi}_v}.$$

It can be seen that $\sum_{u\in\hat{\mathcal{U}}^o}\hat{\phi}_u\log(m_u)$ subject to $\frac{m_u}{m_v} \geq \frac{1}{2}\frac{\hat{\phi}_u}{\hat{\phi}_v}$ and $\sum_{u\in\hat{\mathcal{U}}^o}\hat{\phi}_u = 1$ is maximized when the $\frac{m_u}{\hat{\phi}_u}$ of all users but one is equal to the lower bound given by the constraint, which yields

$$\frac{m_u}{\hat{\phi}_u} = \frac{1}{2}\frac{m_v}{\hat{\phi}_v}, \ \forall u\neq v. \quad (30)$$

This is shown by contradiction. Let us imagine that in the weight allocation that maximizes (29) there exists some other user $u$ for which $\frac{m_u}{\hat{\phi}_u} > \frac{m_v}{2\hat{\phi}_v}$, where $v$ is the user with the largest $m_v/\hat{\phi}_v$ of that allocation. Then, if we increase $m_v$ by $\delta$ and decrease $m_u$ by $\delta$ we have

$$\frac{d}{d\delta}\sum_{u\in\mathcal{U}^o}\hat{\phi}_u\log\left(\frac{\hat{\phi}_u s'_o}{m_u}\right) = -\frac{\hat{\phi}_v}{m_v} + \frac{\hat{\phi}_u}{m_u} > 0$$

and thus (29) increases, which contradicts our assumption that (29) was already maximum. From (30) we have

$$m_u = \frac{\hat{\phi}_u s_o}{\sum_{u'\in\mathcal{U}^o\setminus\{v\}}\hat{\phi}_u + 2\hat{\phi}_v}, \text{ and } m_v = \frac{2\hat{\phi}_v s_o}{\sum_{u'\in\mathcal{U}^o\setminus\{v\}}\hat{\phi}_u + 2\hat{\phi}_v}$$

Combining this with (29) we obtain

$$U^o(\tilde{\mathbf{w}}_o^*) - U^o(\mathbf{w}_o^*) \leq \sum_{u\in\mathcal{U}^o\setminus\{v\}}\hat{\phi}_u\log\left(\sum_{u'\in\mathcal{U}^o\setminus\{v\}}\hat{\phi}_u + 2\hat{\phi}_v\right)$$
$$+ \hat{\phi}_v\log\left(\frac{1}{2}\sum_{u'\in\mathcal{U}^o\setminus\{v\}}\hat{\phi}_u + 2\hat{\phi}_v\right)$$
$$= \log(1+\hat{\phi}_v) + \hat{\phi}_v\log(1/2)$$

If we now compute the $\hat{\phi}_v$ that maximizes this expression we obtain $\hat{\phi}_v = \frac{1}{\log 2} - 1$, and substituting this value

$$U^o(\tilde{\mathbf{w}}_o^*) - U^o(\mathbf{w}_o^*) \leq -\log(\log 2) - \left(\frac{1}{\log 2} - 1\right)\log 2$$

As mentioned at the beginning, the above bounds also applies to $U^o(\tilde{\mathbf{w}}_\mathbf{o}) - U^o(\mathbf{w}_\mathbf{o})$. □

The following lemma proves that the worst case envy is lower bounded by 0.041.

**Lemma 10.** *There exists a game instance for which $U^o(\tilde{\mathbf{w}}_\mathbf{o}) - U^o(\mathbf{w}_\mathbf{o}) = 0.041$.*

*Proof.* Let us consider a scenario with 2 base stations. Let slice $o$ have a share of $s_o$ and one user at each base station with priorities $\phi_1$ and $\phi_2$. Let the loads of the other slices in these two base stations be $a_1 = 1 - s_o - x\phi_2 s_o$ and $a_2 = x\phi_2 s_o$. for a fixed $x > 0$. Let $s_o$ be sufficiently small such that $a_1 > \phi_2 s_o$.

In this setting, the weights of slice $o$ at each station are given by

$$d_1^1 = \frac{s_o\phi_1\frac{a_1}{a_1+d_1^1}}{\phi_1\frac{a_1}{a_1+d_1^1} + \phi_2\frac{a_2}{a_2+d_2^1}}, \text{ and } d_2^1 = \frac{s_o\phi_2\frac{a_2}{a_2+d_2^1}}{\phi_1\frac{a_1}{a_1+d_1^1} + \phi_2\frac{a_2}{a_2+d_2^1}}$$

We distinguish the cases (i) $x \geq 1$ and (ii) $x < 1$.

(i) For $x \geq 1$, we consider slice $o'$ with share $s_{o'} = s_o$ with priorities $\tilde{\phi}_1$ and $\tilde{\phi}_2$, where

$$\frac{\tilde{\phi}_1}{\tilde{\phi}_2} = \frac{\phi_1}{\phi_2}\frac{\frac{a_2-\phi_2 s_o+d_2^1}{a_2+d_2^1}}{\frac{a_1-\phi_1 s_o+d_1^1}{a_1+d_1^1}}$$

We further consider a third slice with only one user in the first base station with $s_3 = a_1 - \phi_1 s_o$ and a fourth slice with a one user in the second base station with $s_4 = a_2 - \phi_2 s_o$. This leads to $d_1^2 = \phi_1 s_o$ and $d_2^2 = \phi_2 s_o$.

If we now let $s_o \to 0$,

$$d_2^1 = \frac{\phi_2 x \phi_2 s_o}{\phi_1(x\phi_2 s_o + d_2^1) + \phi_2 x \phi_2 s_o} s_o = \phi_2 \frac{x \phi_2 s_o}{x\phi_2 s_o + d_2^1} s_o$$

From the above, $d_2^1 = \hat{x}\phi_2 s_o$, where $\hat{x}$ is the unique solution to the equation $x = (x + \hat{x})\hat{x}$. Then, $d_1^1 = s_o - \hat{x}\phi_2 s_o$. From this, we have that in this case

$$U^o(\tilde{\mathbf{w}}) - U^o(\mathbf{w}) = \phi_1 \log\left(\frac{\phi_1 s_o}{s_o - \hat{x}\phi_2 s_o}\right) + \phi_2 \log\left(\frac{\phi_2 s_o}{\hat{x}\phi_2 s_o}\right)$$

$$= \phi_1 \log\left(\frac{\phi_1}{1 - \hat{x} + \hat{x}\phi_1}\right) - (1 - \phi_1)\log(\hat{x})$$

(ii) In case that $x < 1$, we consider slice $o'$ has priorities $\tilde{\phi}_1$ and $\tilde{\phi}_2$, where

$$\frac{\tilde{\phi}_1}{\tilde{\phi}_2} = \frac{\phi_1}{\phi_2} \frac{\frac{a_2 - x\phi_2 s_o + w_2}{a_2 + w_2}}{\frac{a_1 - s_o - x\phi_2 s_o + w_1}{a_1 + w_1}}$$

which leads to $\tilde{w}_1 = (1 - x\phi_2)s_o$ and $\tilde{w}_2 = x\phi_2 s_o$. We further consider a third slice in the first base station with $s_3 = a_1 - (1 - x\phi_2)s_o$. If we now let $s_o \to 0$, we have the same expressions as above for $w_1$ and $w_2$, from which

$$U^o(\tilde{\mathbf{w}}_o) - U^o(\mathbf{w}_o) = \phi_1 \log\left(\frac{s_o - x\phi_2 s_o}{s_o - \hat{x}\phi_2 s_o}\right) + \phi_2 \log\left(\frac{x\phi_2 s_o}{\hat{x}\phi_2 s_o}\right)$$

$$= \phi_1 \log\left(\frac{1 - x + x\phi_1}{1 - \hat{x} + \hat{x}\phi_1}\right) - (1 - \phi_1)\log\left(\frac{x}{\hat{x}}\right)$$

By putting together the cases $x \geq 1$ and $x < 1$, we can obtain a lower bound for the worst-case envy by finding the values of $x$ and $\phi_1$ over $x > 0$ and $\phi_1 \in [0, 1]$ that minimize the following expression

$$\begin{cases} \phi_1 \log\left(\frac{\phi_1}{1-\hat{x}+\hat{x}\phi_1}\right) - (1-\phi_1)\log(\hat{x}), & \text{for } x \geq 1 \\ \phi_1 \log\left(\frac{1-x+x\phi_1}{1-\hat{x}+\hat{x}\phi_1}\right) - (1-\phi_1)\log\left(\frac{x}{\hat{x}}\right), & \text{for } x < 1 \end{cases}$$

By performing the above search numerically, we find a scenario with the following envy level:

$$U^o(\tilde{\mathbf{w}}_o) - U^o(\mathbf{w}_o) = 0.041 \qquad \square$$